\documentclass[chicago, iop, apj]{emulateapj}

\usepackage{natbib}
\usepackage{bm}
\usepackage{times}
\usepackage{graphicx}
\usepackage{color}
\usepackage{amsmath}
\usepackage{amssymb}
\usepackage{fancyvrb}
\usepackage{url}
\usepackage{hyperref}  
\usepackage{verbatim}

\renewcommand{\cite}{\citep}

\newcommand{\nfg}{\mathbf{n}_\textrm{fg}}
\newcommand{\ninst}{\mathbf{n}_\textrm{inst}}

\newcommand{\nfreq}{N_{\nu}}
\newcommand{\npix}{N_{\rm pix}}

\newcommand{\vect}{\boldsymbol}
\newcommand{\mat}{\boldsymbol}

\newcommand{\qed}{\nobreak \ifvmode \relax \else
      \ifdim\lastskip<1.5em \hskip-\lastskip
      \hskip1.5em plus0em minus0.5em \fi \nobreak
      \vrule height0.75em width0.5em depth0.25em\fi}

\begin{document}
\VerbatimFootnotes

%\slugcomment{Submitted to ApJ}

\author{
Eric~R.~Switzer\altaffilmark{1,2} and 
Adrian~Liu\altaffilmark{3,4}
}

\altaffiltext{1}{NASA Goddard Space Flight Center, Greenbelt, MD 20771, USA}
\email{eric.r.switzer@nasa.gov}
\altaffiltext{2}{Canadian Institute for Theoretical Astrophysics,
University of Toronto, 60 St. George St., Toronto, ON, M5S 3H8, Canada}
\altaffiltext{3}{Department of Astronomy, UC Berkeley, Berkeley, CA 94720, USA}
\altaffiltext{4}{Berkeley Center for Cosmological Physics, UC Berkeley, Berkeley, CA 94720, USA}

\shortauthors{Switzer \& Liu}
%\date{\red{April 30, 2014}}  % Arbitrary target date. 

\title{ERASING THE VARIABLE: EMPIRICAL FOREGROUND DISCOVERY FOR GLOBAL $21$\,cm SPECTRUM EXPERIMENTS
\shorttitle{Foreground discovery for the $21$\,cm monopole}}

\begin{abstract}
Spectral measurements of the $21$\,cm monopole background have the promise of revealing the bulk energetic properties and ionization state of our universe from $z\sim 6-30$. Synchrotron foregrounds are orders of magnitude larger than the cosmological signal, and are the principal challenge faced by these experiments. While synchrotron radiation is thought to be spectrally smooth and described by relatively few degrees of freedom, the instrumental response to bright foregrounds may be much more complex. To deal with such complexities, we develop an approach that discovers contaminated spectral modes using spatial fluctuations of the measured data. This approach exploits the fact that foregrounds vary across the sky while the signal does not. The discovered modes are projected out of each line of sight of a data cube. An angular weighting then optimizes the cosmological signal amplitude estimate by giving preference to lower-noise regions. Using this method, we show that it is essential for the passband to be stable to at least $\sim 10^{-4}$. In contrast, the constraints on the spectral smoothness of the absolute calibration are mainly aesthetic if one is able to take advantage of spatial information. To the extent it is understood, controlling polarization to intensity leakage at the $\sim 10^{-2}$ level will also be essential to rejecting Faraday rotation of the polarized synchrotron emission.
\end{abstract}

\maketitle

\section{Introduction}

One of the richest yet least understood narratives in cosmology is the formation of the complex structure that we see today out of the simple initial conditions implied by the cosmic microwave background (CMB). The first luminous objects are thought to have formed at $z\sim 20-30$ through collapse in $10^6-10^8\,{\rm M}_\odot$ halos \citep{2001PhR...349..125B, 2013RPPh...76k2901B}. The radiation from these objects heated and then reionized the intergalactic medium (IGM). There are several sources of complementary information about the evolution of ionization in this epoch. 

The CMB temperature anisotropy is damped by the total Thomson depth to free electrons. The Planck collaboration has used this effect, combined with a constraint on the scalar amplitude from gravitational lensing, to measure the optical depth through reionization \citep{2013arXiv1303.5076P}. In addition, Thomson scattering through the reionization epoch generates a unique polarization signature on large angular scales. WMAP has measured the total optical depth using this polarization signature \citep{2013ApJS..208...20B}. These are integral constraints on the free electron abundance and can be translated into a central reionization redshift of $10.6 \pm 1.1$ \citep{2013ApJS..208...20B}.

Once the IGM is highly ionized, it is transparent to Ly-$\alpha$ photons. Absorption measurements along sight lines to high-redshift quasars indicate that reionization must have ended by $z<6$ \citep{2006AJ....132..117F}. Absorption saturates at low abundance, so these should be taken as bounds on the end of reionization, which could still have been largely complete at redshifts higher than $6$. 

Recently, two methods have been developed to place much more direct bounds on the duration of reionization. The ionization process is thought to be spatially patchy, as local sources of radiation blow ionized bubbles that coalesce into the fully reionized IGM. CMB photons scattering from this patchy screen produce an additional kinetic Sunyaev-Zel'dovich anisotropy appearing most clearly at $\ell > 3000$, where the primary CMB is negligible \citep{1998PhRvL..81.2004K}. Upper limits on this effect translate into a model-dependent upper bound on the duration of reionization \cite{2012ApJ...756...65Z} and hold the promise of direct detection of patchy structure in the near future.

The patchy structure of reionization can also be observed directly in three dimensions using emission of neutral hydrogen through its $21$\,cm line \citep{2006PhR...433..181F}. Recent bounds from GMRT \citep{2013MNRAS.433..639P}, MWA \citep{2014PhRvD..89b3002D}, and PAPER \citep{2014ApJ...788..106P, 2013ApJ...768L..36P} are marching down to the expected level of fluctuations, in parallel to efforts at LOFAR \citep{2013A&A...556A...2V}. An alternative to measuring the $21$\,cm anisotropy is to measure the signal of its global emission (or absorption at earlier times) \citep{2008PhRvD..78j3511P, 2010PhRvD..82b3006P}, which reveals the bulk energetic properties and ionization state of the universe during reionization and preceding epochs when the first luminous structures were forming. Global $21$\,cm experiments include EDGES \citep{2010Natur.468..796B} and SCI-HI \cite{2014ApJ...782L...9V} (which have both reported bounds), LEDA\footnote{{\tt http://www.cfa.harvard.edu/LEDA/science.html}} and the proposed DARE mission \citep{2012AdSpR..49..433B}.  

The frequencies of interest in these global studies are $\sim 50-200$\,MHz, and fiducial theoretical models suggest a maximum contrast of $\sim 100$\,mK relative to the synchrotron emission of the galaxy, which can vary $\sim 10^2-10^5$~K across the sky and frequency range. Astrophysical synchrotron emission is thought to be fully described by a handful of spectrally smooth functions that can be distinguished from the variation of the global reionization signal. 

Extremely bright foregrounds make the measurement susceptible to instrumental systematics. For example, if an instrument with a $1\%$ perturbation to the spectral calibration observed a $500$~K power law, subtraction of a pure power law would leave a $5$~K residual, significantly larger than the signal. Through careful instrumental design, the level of instrumental systematics may be controlled but generally cannot be nulled entirely. 

Here, we develop methods that can be used to constrain the cosmological signal in these heavily contaminated data. Following the monopole nature of the signal, experiments to date have mapped the sky with very large beams \citep{2010Natur.468..796B, 2014ApJ...782L...9V}. However, a unique trait of the foregrounds is that they vary across the sky, while the signal is constant. This regime is the opposite of the situation normally found in analysis of small signals, where a modulated signal is pulled out of foregrounds. \citet{2013PhRvD..87d3002L} (henceforth L13) proposed that experiments seeking to measure the global signal should also resolve the sky with an instrumental beam. This allows selective weighting against regions of high contamination and allows for the use of angular correlation information to reject foregrounds. The additional spatial resolution yields higher fidelity recovery of the cosmological $21$\,cm spectrum.

Here, we extend the ideas in L13 to a method that uses the spatial fluctuations of foregrounds in the data to discover contaminated spectral modes. A similar idea has been employed successfully in $21$\,cm large-scale structure measurements \cite{2013MNRAS.434L..46S, 2013ApJ...763L..20M, 2010Natur.466..463C} at $z\sim 1$ and has been suggested for cleaning ionospheric contamination \citep{2014MNRAS.437.1056V}.

Discovery of foreground spectral modes in the measured data makes the method more robust to assumptions about the foreground contamination. For example, now if the instrumental passband has a $1\%$ ripple, the largest foreground mode discovered in our foreground cleaning method will also self-consistently exhibit this ripple. Generally, instrumental systematics take relatively clean and smooth functions of frequency from synchrotron emission and convert them into a more complex structure that requires additional spectral functions to describe. We argue that the primary goal in instrumental design should be to prevent proliferation of bright, new foreground modes in the data. Each new foreground degree of freedom produced by instrumental response to foregrounds results in more signal loss and makes discovery of the signal more ambiguous. 

The methods described here of (1) using spatial variation to discover spectral foreground modes, which can then be projected out and (2) down-weighting known spatial areas of high contamination (the galaxy) provide the strongest methods for recovering the global $21$\,cm signal in the absence of additional prior information about the foregrounds or instrumental response. While the algorithm of mode subtraction and angular weighting is intuitive, we develop it from the ground up to expose several implicit choices and possible pitfalls. 

Recently, \citet{2014arXiv1404.0887B} argued that a dipole gain pattern can be calibrated in an interferometric array. However, additional variations in spectral response due to factors such as the analog-to-digital converter, reflection, and signal loss after the antenna were not included. Our goal here is to understand how data analysis can be made more robust to this class of instrumental response (or any other source of foreground covariance), or alternately how tightly certain instrumental tolerances must be constrained.

In Section~\ref{sec:globsig} we review the basic properties of the global signal and describe our foreground model. Section~\ref{sec:simplifiedmodel} builds up the estimator for joint foreground and signal estimation using a simplified model of spectra along independent sight lines. Section~\ref{sec:instrum} considers implications of this model for passband calibration. Section~\ref{sec:spatial_information} develops spatial weights, and Section~\ref{sec:separable_cov} combines the estimators with spatial and spectral weights. Section~\ref{sec:analysis_considerations} describes a number of considerations for using the methods developed here and for global $21$\,cm signal estimation in general. We discuss telescope beam width, the foreground monopole, how aggressively foreground modes should be removed, and susceptibility to Faraday rotation. We also consider mode removal of the pre-reionization absorption feature and extensions of the simple template amplitude constraint considered throughout. We summarize our conclusions in Section~\ref{sec:discussion}.

\section{Global Signal and Foreground Models}
\label{sec:globsig}

From radiative transport arguments, the brightness temperature of $21$\,cm radiation is
\begin{equation}
T_b \approx 27 (1-\bar x_i) \left ( \frac{T_S - T_{\rm CMB}}{T_S} \right ) \left ( \frac{1+z}{10} \right )^{1/2}\,{\rm mK}, 
\end{equation}
where $\bar x_i$ is the mean ionization fraction, $T_S$ is the spin temperature of the hyperfine transition, and $T_{\rm CMB}$ is the CMB temperature. The basic physics of the spin-temperature coupling is well understood \cite{2010PhRvD..82b3006P}, but the detailed astrophysical processes that determine the coupling strength and the gas temperature are still conjectural.

For $z> 200$, the universe is dense enough that electron collisional interactions drive the spin temperature to the gas temperature, which is cooling faster than the CMB. This produces absorption. By $z \sim 30$, the universe is sufficiently rarefied that the spin temperature is better coupled to the CMB bath, and absorption is expected to subside. Once the first luminous objects form, these produce radiation that drives the Wouthuysen-Field \citep{1952Phy....18...75W, 1958PIRE...46..240F} coupling of $T_S$ again to the gas temperature, leading to a second absorption feature. Then, X-ray heating of the gas can drive the spin temperature above the CMB temperature, leading to emission. As these luminous processes proceed and increase, they also ionize the IGM, which causes the signal to disappear as $1-\bar x_i \rightarrow 0$ \citep{2010PhRvD..82b3006P}. 

We will find it convenient to have a reionization model with a small number of parameters, rather than a full spectrum. For concreteness, we will spend most of the paper focusing on the evolution of the brightness temperature as the universe is reionized, rather than the absorption dip as heating begins. However, our methods are applicable at all redshifts, and in Section~\ref{sec:DarkAges} we will briefly discuss the pre-reionization dip.

\begin{figure}
\epsscale{1.2}
\plotone{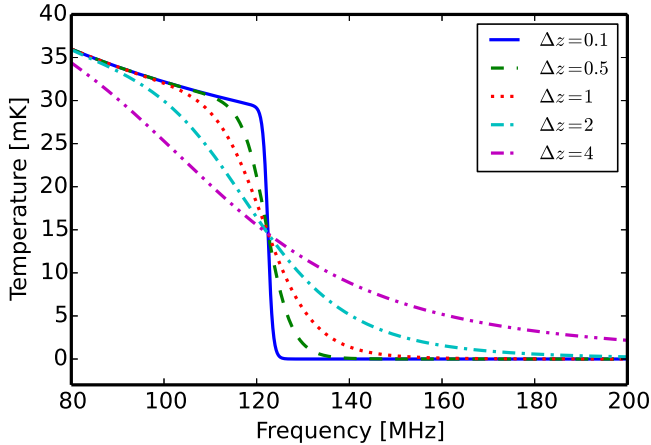}
\caption{Brightness temperature of the global $21$\,cm emission in a simplified two-parameter model of reionization. The central redshift is taken from the WMAP9 constraint at $z=10.6$, and several representative values for the duration of reionization are shown.\label{fig:tanh_model}}
\end{figure}

At the beginning of the reionization epoch, it is thought that the spin temperature is strongly coupled to the gas and that the gas is heated, driving $21$\,cm emission. As the ionization proceeds, this emission dies away. A simple way of parameterizing this is \citep{2010PhRvD..82b3006P} 
\begin{equation}
\bar x_i(z) = \frac{1}{2} \left [ 1 + \tanh \left ( \frac{z_r - z}{\Delta z} \right ) \right ],
\end{equation}
so that the brightness temperature scales as (Figure~\ref{fig:tanh_model})
\begin{equation}
T_b(z) = \frac{27}{2} \left ( \frac{1+z}{10} \right )^{1/2} \left [ 1 + \tanh \left ( \frac{z_r - z}{\Delta z} \right ) \right ]\,{\rm mK}
\label{eqn:tanh_model}
\end{equation}
where we have assumed that $T_S \gg T_{\rm CMB}$. 

\begin{figure}
\epsscale{1.2}
\plotone{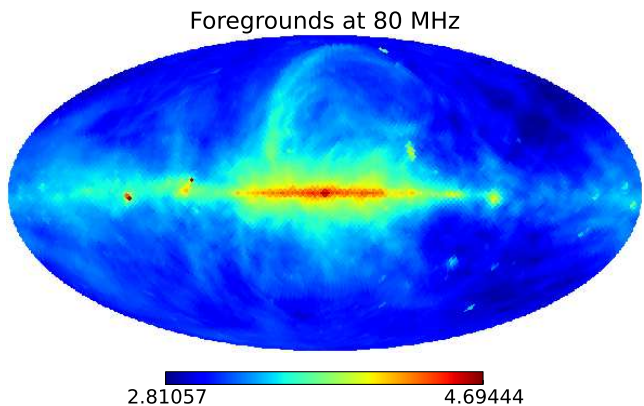}
\caption{Foreground model at 80 MHz, displayed as $\log(T)$.}
\label{fig:fg_model}
\end{figure}

Our model for the diffuse intensity (Stokes-$I$) of the foregrounds is based on an extended version of the Global Sky Model (GSM) of \citet{2008MNRAS.388..247D} that is developed in L13. The original GSM used a principal component analysis to extrapolate and interpolate between previous galactic surveys at a wide range of frequencies. It is based on three spectral components and is accurate down to $\approx 10^\circ$. L13 extends this model by adding additional (less smooth) spectral eigenmodes as well as a population of power-law sources with randomly distributed spectral indices, drawn from the $dN/dS$ brightness distribution of \citet{2002ApJ...564..576D}. The goal of adding these components is to boost the rank of the foreground covariance to reflect spectral fluctuations that can be expected on the real sky. These cannot be probed in the original rank-3 GSM.  An example map of our foreground model at $80\,\textrm{MHz}$ is shown in Figure~\ref{fig:fg_model}.

In Section~\ref{ssec:faraday}, we will additionally consider the impact of Faraday rotation and polarized foregrounds.  Because this is somewhat peripheral to the development of our data analysis methods in the next few sections, we will defer our discussion of our foreground polarization model until then.

\section{Global $21$\,cm Observations Along Independent Lines of Sight}
\label{sec:simplifiedmodel}

A global $21$\,cm experiment that maps the sky will have a finite number of resolution elements across its field, defined by the beam. In each one of these resolution elements, it observes the spectrum of incoming radiation. A reasonable approximation is to split the survey into $N_\theta$ spectra of length $N_\nu$ along all of the angular resolution elements. Let $\vect{y}_i$ be that vector, where $i$ indexes the sight line ($1$ to $N_\theta$) and the length of the vector is $N_\nu$. In this section, we will build up infrastructure and intuition using this simplified case of $N_\theta$ independent and identically distributed sight lines. Section~\ref{sec:separable_cov} extends this to a method that treats realistic foregrounds with proper angular correlations.  Because angular correlations become largely irrelevant when considering wide-beam experiments with essentially no angular sensitivity, much of the intuition developed here is directly transferable to the analysis of such experiments.

In writing an estimator, there are many choices for aspects of the global signal that could be estimated. The goal could be to constrain (1) an arbitrary spectrum of the cosmological $21$\,cm evolution, (2) some modes of its variation that relate to physical parameters, (3) some amplitudes that are based on external information about the expected redshift of reionization, or (4) the amplitude of a template of a provisional global $21$\,cm signal. For our purposes here, it is simplest to develop estimators for the amplitude of an assumed $21$\,cm signal template. Initially, experiments will simply be stepping down in bounds and seeking some evidence of $z \sim 20$ absorption and heating or a $z\sim 11$ reionization signal---an amplitude constraint could provide the simplest, clear indication of a cosmological signal. Section~\ref{ssec:extended_template} considers other estimation regimes in light of the methods developed here. 

Let the assumed template of the global signal be a vector $\vect{x}$ of length $N_\nu$ and normalized to have a maximum of $1$. Multiply by some amplitude $\alpha$ to get the $21$\,cm signal. Then the observed spectrum is the sum of signal, thermal noise $\vect{n}_{{\rm inst}, i}$, and foregrounds $\vect{n}_{{\rm fg}, i}$:
\begin{equation}
\vect{y}_i = \alpha \vect{x} + \vect{n}_{{\rm fg}, i} + \vect{n}_{{\rm inst}, i}.
\label{eq:iid_datamodel}
\end{equation}
Initially we will assume that the foregrounds and noise are identically and independently normally distributed along each line of sight $i$, $\vect{n}_{{\rm fg}, i} \sim N(0, \mathbf{\Sigma}_{\rm fg})$ and $\vect{n}_{{\rm inst}, i} \sim N(0, \mathbf{\Sigma}_{\rm inst})$. In reality, the foreground field is strongly correlated in angle and non-Gaussian. These issues will be examined in subsequent sections. In contrast, thermal noise is uncorrelated between sight lines and normally distributed: $\vect{n}_{{\rm inst}, i} \sim N(0, \mathbf{\Sigma}_{\rm inst})$ will remain an excellent approximation. Throughout, the $N_\nu \times N_\nu$ matrix $\mathbf{\Sigma} \equiv  \mathbf{\Sigma}_{\rm inst} + \mathbf{\Sigma}_{\rm fg}$ will refer to the total $(\nu, \nu')$ covariance.

\subsection{Known Covariance}

If the foregrounds and thermal noise are drawn from a total, known covariance $\mathbf{\Sigma}$, then the maximum likelihood estimate for the template amplitude is 
\begin{equation}
\hat \alpha_{\rm ML} = (\vect{x}^T \mathbf{\Sigma}^{-1} \vect{x})^{-1} \vect{x}^T \mathbf{\Sigma}^{-1} \vect{\bar y}
\label{eqn:known_cov_est}
\end{equation}
where 
\begin{equation}
\label{eq:StraightAverage}
\vect{\bar y} = N_\theta^{-1} \sum_{i=1}^{N_\theta} \vect{y}_i
\end{equation}
is the mean spectrum along all lines of sight.

This can be understood as ``deweighting'' the foregrounds ($\mathbf{\Sigma}^{-1} \vect{\bar y}$), projecting onto the signal template ($\vect{x}^T \mathbf{\Sigma}^{-1} \vect{\bar y}$), and then applying a normalization to account for the weights. The estimated amplitude is normally distributed, and its error is 
\begin{equation}
\label{eq:UsualErrorBars}
{\rm Var}(\hat \alpha_{\rm ML}) |_{\mathbf{\Sigma}} = (\vect{x}^T \mathbf{\Sigma}^{-1} \vect{x})^{-1}.
\end{equation}

To convert this into a more intuitive cleaning process, take the eigendecomposition of the covariance $\mathbf{\Sigma} = \mat{V} \mathbf{\Lambda} \mat{V}^T$. The cleaning operation $\mathbf{\Sigma}^{-1} \vect{\bar y}$ is then $\mat{V} \mathbf{\Lambda}^{-1} \mat{V}^T \vect{\bar y}$. Here $\mat{V}^T$ projects the data onto a basis where the contaminated modes $\vect{v}_j$ (the $j$'th row of $\mat{V}$) are orthonormal. Those modes are weighted against $\lambda_j^{-1} = [\mathbf{\Lambda}^{-1}]_{jj}$, where the highest variance, most contaminated modes are down-weighted. Then, $\mat{V}$ projects this down-weighted basis back onto the spectral basis.

In the usual assumptions, the foregrounds are not full rank, even if the spectral covariance $\boldsymbol \Sigma$---which contains instrumental noise---is full rank. Often, the foregrounds are instead described as some set of contaminated {\it modes} $\vect{v}_j$ where $j$ ranges from one to the number of contaminated modes $N_{\rm fg}$, or
\begin{equation}
\vect{n}_{{\rm fg},i} = \sum_j^{N_{\rm fg}} a_{i,j} \vect{v}_j.
\end{equation}

If the foreground covariance is described by only $N_{\rm fg}$ highly contaminated modes, a robust cleaning method is to null those modes entirely, setting $\lambda \rightarrow \infty$ artificially \citep{2012MNRAS.419.3491L}. This is equivalent to fitting and subtracting those spectral modes from each line of sight, as 
\begin{equation}
\vect{\bar y}_{\rm clean} = \sum_j^{N_{\rm fg}} (1 - \vect{v}_j \vect{v}_j^T) \vect{\bar y},
\label{eq:modeclean}
\end{equation}
where the $\vect{v}_i$ are normalized so that $\vect{v}^T_i \vect{v}_i = 1$.

Here we have assumed that the foreground spectral functions are known in advance and can be projected out. This is essentially the same as arguments for subtracting polynomials or power laws along the line of sight. A crucial difference is that there is no assumption of smoothness---if the foregrounds were known to have a particular spectral shape, that could be represented in the vectors $\vect{v}_i$. The essential property of the foregrounds that allows for their removal is not that they are smooth, but that they are described by few functions. In the limit that the number of orthogonal functions removed approaches $N_\nu$, all of the signal is removed because $\{ \vect{v}_i, i = 1\ldots N_\nu \}$ spans the space, assuming orthogonality of the $\vect{v}_i$. The rank of bright foregrounds is the primary determinant of cosmological signal that can be extracted.

\begin{figure}
\epsscale{1.2}
\plotone{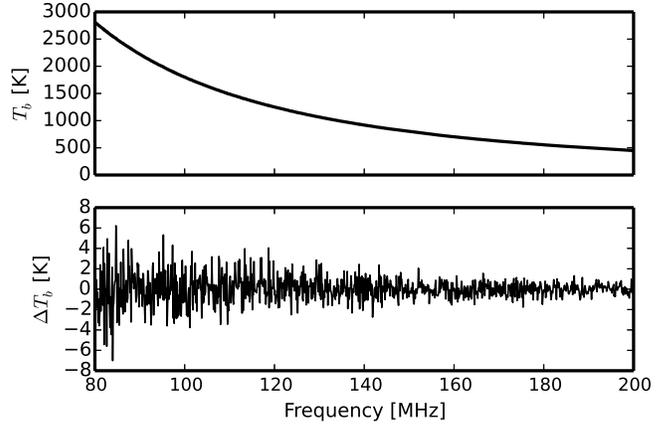}
\caption{Impact of calibration uncertainty. {\it Top:} a power-law foreground spectrum similar to the one observed in \citet{2014ApJ...782L...9V}, multiplied by a $0.1\%$ calibration error. {\it Bottom:} residuals when a smooth power-law spectrum is subtracted from data with a calibration error. For contrast, the global signal at these redshifts is expected to be $\sim 30$\,mK. Also note that the calibration error will not integrate down thermally.
\label{fig:smoothness_demo}}
\end{figure}

Subtraction of bright foregrounds has proven to be significantly more challenging in practice due to instrumental effects \citep{2013MNRAS.434L..46S}. Figure~\ref{fig:smoothness_demo} shows an example of $0.1\%$ calibration for a power-law foreground spectrum similar in amplitude to the one reported by \citet{2014ApJ...782L...9V} and scaled to the slightly different frequencies of interest here. This represents the case where the $\vect{v}_i$ mode removed assumes a pure power law but the actual measurement reflects the instrument's response to $\vect{v}_i$. Residuals are considerably larger than the signal.

One approach to treating the passband is to control it through instrumental design and to measure it very precisely prior to the experiment. In principle, this could be achieved by long integrations on a calibration reference that is known to be smooth. This carries the challenge of calibrating on a source other than the sky. An alternative is to recognize that the foregrounds are a bright spectral reference, viewed through the same survey program as the primary science. One calibration scheme would then consist of integrating down on bright synchrotron emission and renormalizing the spectrum under the assumption that the emission is spectrally smooth \citep{2014ApJ...782L...9V}. 

An alternative to calibrating the instrument to meet the smooth mode functions $\{ \vect{v}_i \}$ is to adjust the mode functions to reflect imperfections in the instrument. In the next section we develop a method for discovering foreground spectral mode functions in the data based on their {\it spatial} variations. The central idea is that the cosmological signal is a monopole spectrum, so the spectrum of anything that fluctuates across the sky must reflect some foreground modes (hence the title ``erasing the variable"). If there are monopole foreground components, they are formally indistinguishable from the cosmological $21$\,cm monopole in this picture (in the absence of prior information). 

Measurement of $\{ \vect{v}_i \}$ modes within the data may also discover unanticipated or unconstrained instrumental response systematics. Examples include (1) a passband that varies with time or pointing, which causes a response to synchrotron radiation that varies across the sky, (2) polarization leakage, which can cause Faraday (spectral) rotation that varies across the sky, (3) terrestrial interference that varies with instrument pointing, and (4) unmodeled frequency dependence of the beam. Even if the foregrounds were intrinsically a simple rank-1 spectral function, the instrument response will tend to proliferate the rank of the foreground covariance. This limits the ultimate sensitivity to the global signal. While many of these sources of systematics may be tightly controlled by the instrument's construction, residual imperfections will not be known in advance and could be discovered by this method.

We will assume throughout that the true synchrotron foregrounds are described by a small number of modes $N_{\rm fg, true}$. The instrument observes these bright foregrounds and records spectral data that may need some $N_{\rm fg} \geq N_{\rm fg, true}$ functions to describe the contamination.

\subsection{Covariance from the Measurements}
\label{sec:cov_from_meas}

This section considers the joint determination of the global signal amplitude $\alpha$ and the contaminated foreground modes. The statistical methods described and extended here were developed by \citet{rao1967}. The central idea in these methods is to begin with a poor, contaminated estimator for the $21$\,cm signal amplitude. This initial estimator can then be cleaned up by projecting out correlations with the foregrounds. 

To reflect the reality that often one does not know the covariance a priori, first consider a simple estimator that does not use any information from the spectral covariance.  This will serve as our starting point for constructing an estimator that simultaneously estimates the spectral covariance from the data. Again, let $\vect{x}$ be the spectral template of the cosmological signal that we would like to constrain. The simplest (and arguably worst) estimator we could develop for the amplitude is 
\begin{equation}
T(\parallel \vect{x}) = (\vect{x}^T \vect{x})^{-1} \vect{x}^T \vect{\bar y},
\end{equation}
which takes the dot product of the $21$\,cm template and observed data ($\vect{x}^T \vect{\bar y}$) and normalizes through $(\vect{x}^T \vect{x})^{-1}$. We refer to this as $T(\parallel \vect{x})$ because it estimates the amplitude of the component of $\vect{\bar y}$ parallel to the signal $\vect{x}$. This is exactly the estimator that would be used if there were no frequency correlations in the noise. What makes this a very poor estimator is that it is strongly correlated with the foreground modes and will therefore contain strong foreground residuals. In contrast, Equation~\eqref{eqn:known_cov_est} in the previous section used the known frequency covariance to deweight the foreground contamination.

We can begin to improve upon our simple estimator by constructing spectral modes that contain no cosmological signal by design.  Using a Gram-Schmidt process, we can make a set of vectors $\vect{z}_i$ that span the space orthogonal to the theoretical template of the cosmological signal $\vect{x}$, so that $\vect{z}_i^T \vect{x}=0$. There are $N_\nu -1$ such vectors, and the choice of these vectors is completely arbitrary at this point. Any spectrum can be represented as a sum
\begin{equation}
\label{eq:AlphaXBetaZ}
\vect{y} = \alpha \vect{x} + \sum_i \beta_i \vect{z}_i
\end{equation}
because $\{ \vect{x}, \vect{z}_i \}$ span the $N_\nu$-dimensional space. We can pack these vectors $\vect{z}_i$ into a matrix $\mat{Z}$ and write an estimator for all of the spectral information orthogonal to $\vect{x}$, $T(\perp \vect{x}) = \mat{Z}^T \vect{\bar y}$. This represents all of the components of $\vect{\bar y}$ orthonormal to the signal and is dominated by foregrounds. (Note that we could choose to normalize this for general vectors $\vect{z}$ and find $(\mat{Z}^T \mat{Z})^{-1} \mat{Z}^T \vect{\bar y}$. This would make the equations more complex but would not modify the estimator for the cosmological quantity.)

The covariance between $T(\parallel \vect{x})$ and $T(\perp \vect{x})$ is 
\begin{eqnarray}
{\rm Cov}_{\parallel, \perp} &=& \left( \begin{array}{cc} \mat{C}_{\parallel, \parallel} & \mat{C}_{\parallel, \perp} \label{eq:ParaPerpCov} \\ 
\mat{C}_{\perp, \parallel}  & \mat{C}_{\perp, \perp} \end{array} \right) \\
&=& \left( \begin{array}{cc} (\vect{x}^T \vect{x})^{-1} \vect{x}^T \mathbf{\Sigma} \vect{x} (\vect{x}^T \vect{x})^{-1} & (\vect{x}^T \vect{x})^{-1} \vect{x}^T \mathbf{\Sigma} \mat{Z} \\ \mat{Z}^T \mathbf{\Sigma} \vect{x} (\vect{x}^T \vect{x})^{-1} & \mat{Z}^T \mathbf{\Sigma} \mat{Z} \end{array} \right). \nonumber
\end{eqnarray}
The off-diagonal terms $(\vect{x}^T \vect{x})^{-1} \vect{x}^T \mathbf{\Sigma} \mat{Z}$ represent correlations between $T(\parallel \vect{x})$ and $T(\perp \vect{x})$ caused by the $(\nu, \nu')$ covariance of foregrounds in $\mathbf{\Sigma}$. If the noise terms were only thermal fluctuations, $\mathbf{\Sigma} = \sigma_{\rm inst}^2 \mathbf{1}$, and by the construction $\vect{x}^T \mat{Z} = 0$, the off-diagonal terms vanish. In other words, an estimator $T(\perp \vect{x})$ that is known to contain only contaminants (and no cosmological signal) is correlated with our estimate of the cosmological signal, so the latter must be contaminated.

With knowledge of spectral correlations, we can form an improved, adjusted estimator
\begin{equation}
\hat \alpha = T(\parallel \vect{x}) - \mat{C}_{\parallel, \perp} \mat{C}_{\perp, \perp}^{-1} T(\perp \vect{x}) \label{eqn:adjusted}
\end{equation}
that projects foreground frequency correlations out of the $T(\parallel\,\vect{x})$ estimator. Intuitively, this estimator instructs us to make an estimate $T(\perp \vect{x}) $ of the portion of the measurement that is known to contain only foregrounds.  Off-diagonal elements of the covariance matrix then allow the level of foreground leakage (into our estimate $T(\parallel \vect{x})$ of the cosmological signal) to be predicted and subtracted off.  A similar technique was used recently in a power spectrum analysis of PAPER data \citep{2014ApJ...788..106P}.

So far, we have assumed that $\mathbf{\Sigma}$ is known, but the same covariance adjustment can be performed with respect to an estimated $\mathbf{\hat \Sigma}$. Without perfect knowledge of $\mathbf{\Sigma}$, the error bars grow but the estimator remains unbiased. The $(\nu, \nu')$ covariance can be estimated empirically using
\begin{equation}
\mathbf{\hat \Sigma} = (N_\theta-1)^{-1} \sum_{i=1}^{N_\theta} (\vect{y}_i - \vect{\bar y})(\vect{y}_i - \vect{\bar y})^T
\label{eqn:sample_cov}
\end{equation}

Writing out Equation~\eqref{eqn:adjusted},
\begin{equation}
\hat \alpha = (\vect{x}^T \vect{x})^{-1} \vect{x}^T (1 - \mathbf{\Pi}) \vect{\bar y}
\end{equation}
where $\mathbf{\Pi} = \mathbf{\hat \Sigma} \mat{Z} (\mat{Z}^T \mathbf{\hat \Sigma} \mat{Z})^{-1} \mat{Z}^T$. Remarkably, under our current assumptions where $\mat{Z}$ spans the rest of the $N_\nu$ dimensions, this estimator is equivalent to (see Appendix~\ref{app:lemma_7b})
\begin{equation}
\hat \alpha = (\vect{x}^T \mathbf{\hat \Sigma}^{-1} \vect{x})^{-1} \vect{x}^T \mathbf{\hat \Sigma}^{-1} \vect{\bar y},
\label{iid_full_rank_est}
\end{equation}
which is precisely the same as Equation~\eqref{eqn:known_cov_est}, except with the estimated spectral covariance $\mathbf{\hat \Sigma}$ replacing the known covariance $\mathbf{ \Sigma}$.

\citet{gleser1972} develop the same result through a completely independent method. They write down the joint likelihood of the covariance and mean (cosmological signal), where the covariance is Wishart-distributed and the mean is normally distributed. They then maximize the likelihood and find the same result, showing that the choice $\mathbf{\Sigma} \rightarrow \mathbf{\hat \Sigma}$ coincides with the maximum likelihood.

In summary, the proposed procedure is to
\begin{itemize}
\item Find the sample mean and $(\nu, \nu')$ covariance across observed lines of sight.
\item Invert the measured covariance and use it to down-weight contaminated modes in the data.
\item Find the inner product of the cleaned data and the signal template, then normalize.
\end{itemize}

While the maximum of the likelihood has the same form as the case where the covariance is known, the distribution of the estimated $21$\,cm global signal template amplitude is no longer Gaussian, and is generally broader. These changes are due to the fact that the data are also used for the covariance estimation, which uses up degrees of freedom in determining the foreground modes. \citet{rao1967} shows that 
\begin{equation}
\label{eq:RaoCovariance}
{\rm Var}(\hat \alpha) = \frac{N_\theta - 1}{N_\theta - 1 - r} [\mat{C}_{\parallel, \parallel} - \mat{C}_{\parallel, \perp} \mat{C}_{\perp, \perp}^{-1} \mat{C}_{\perp, \parallel}],
\end{equation}
where $r$ is the rank of the additional degrees of freedom that are estimated. In our case here, $r=N_\nu - 1$ ($\mat{Z}$ spans the rest of the spectral space) so that  
\begin{equation}
{\rm Var}(\hat \alpha) = \left ( 1 + \frac{N_\nu - 1}{N_\theta - N_\nu} \right ) (\vect{x}^T \mathbf{\hat \Sigma}^{-1} \vect{x})^{-1},
\end{equation}
where we have also plugged in the explicit forms for $\mat{C}_{\parallel, \parallel}$, $ \mat{C}_{\parallel, \perp}$,$ \mat{C}_{\perp, \perp}$, and $ \mat{C}_{\perp, \parallel}$ from Equation~\eqref{eq:ParaPerpCov}.  We see that the covariance is enhanced relative to the case of perfect foreground covariance knowledge (Equation~\eqref{eq:UsualErrorBars}) by a factor related to the number of frequencies and independent lines of sight. The error diverges when $N_\theta$ approaches $N_\nu$ from above. The rank $r(\mathbf{\hat \Sigma}) \leq {\rm min}(N_\theta, N_\nu)$, and for $N_\theta \gg N_\nu$, the covariance is well measured. This suggests that the optimal limit is to have many more resolution elements than spectral bins. 

There is little instrumental limitation on the number of spectral bins $N_\nu$ over the bands observed. In contrast, there is a hard limit on the number of ``independent'' lines of sight that can be observed, $N_\theta$. At the frequencies of interest for the global $21$\,cm signal, the beam size tends to be large by diffraction. Angular correlations and noise generally limit the number of independent samples of the spectrum (see Section~\ref{sec:res_elt}).

The requirement $N_\theta > N_\nu$ stems from the fact that $\mathbf{\hat \Sigma}$ is trying to estimate a full $N_\nu$ rank covariance. In general, the sensitivity to $\alpha$ should not depend on the number of spectral bins in the survey, once they become fine enough to resolve the signal. The solution is to instead estimate some number $N_{\rm fg}$ of contaminated spectral functions, less than the number of spectral bins. Then the estimation becomes independent of the $N_\nu$, so long as $N_\nu > N_{\rm fg}$.

\subsection{Empirically Determined Foregrounds of Limited Rank}
\label{sec:limited_rank_iid}

The key step in the covariance adjustment scheme of the previous section was the formation and application of $\mathbf{\Pi} = \mathbf{\hat \Sigma} \mat{Z} (\mat{Z}^T \mathbf{\hat \Sigma} \mat{Z})^{-1} \mat{Z}^T$, which modeled contaminated modes in the original ``poor" estimator for later subtraction.  The modes formed a basis $\mat{Z}$ orthogonal to the signal $\vect{x}$.  The specific choice of vectors in $\mat{Z}$ was arbitrary so long as they were orthonormal (for simplicity) and spanned the spectral subspace orthogonal to the signal.  For concreteness, we suggested forming this basis blindly using a Gram-Schmidt process.  While formally a solution to our problem, this is not a particularly efficient way to implement our recipe in practice, for each of the resulting basis vectors will be a linear combination of noise and foreground modes.  On the other hand, the foregrounds should be describable by a small number of modes \citep{2012MNRAS.419.3491L,2014arXiv1404.0887B}.  A sensible alternative would therefore be to intelligently partition the basis $\mat{Z}$ into two sub-bases, one that is of a relatively low rank $N_{\rm fg}$ containing the foregrounds and another of rank $N_\nu - N_{\rm fg} -1$ consisting of the remaining noise-dominated modes.  Computationally, this means that rather than estimating the full $N_\nu$ rank covariance from the data, we only need to estimate a subset of contaminated modes and their amplitudes.  This limits the number of degrees of freedom $r$ that need to be estimated from the data, which as we saw from Equation~\eqref{eq:RaoCovariance} is crucial for keeping the final error bars small.

In the rank-restricted foreground approximation, the data along a line of sight are the cosmological signal plus some amplitudes $\vect{\beta}$ times foreground modes $\mat{F}$ plus thermal noise, as 
\begin{equation}
\vect{y} = \alpha \vect{x} + \mat{F} \vect{\beta} + \vect{n}_{t} \Rightarrow \mathbf{\Sigma} = \mat{F} \mathbf{\Gamma} \mat{F}^T + \sigma_{\rm inst}^2 \mat{1}, \label{eqn:limrankmodel}
\end{equation}
where we have taken the thermal noise to be stationary for simplicity and let $\mathbf{\Gamma}$ be the eigenvalue matrix of the foreground spectral covariance $\mat{F} \langle \vect{\beta} \vect{\beta}^T \rangle \mat{F}^T$, quantifying the power of the foreground modes $\mat{F}$. The amplitude-ordered eigenvalue spectrum of the total covariance $\mathbf{\Sigma}$ in this case would show some large contaminant amplitudes followed by a noise floor set by $\sigma$. 

Unlike $\mat{Z}$ of the previous section, the foreground mode vectors in $\mat{F}$ alone do not span the entire spectral subspace orthogonal to the cosmological signal.  The remaining portion of the space is spanned by basis vectors that describe instrumental noise, under the assumptions of Equation~\eqref{eqn:limrankmodel}. These vectors can be formed, like before, using a Gram-Schmidt process (this time relative to the signal $\vect{x}$ and foregrounds $\mat{F}$) and packed into a matrix $\mat{G}$. The observed spectrum $\vect{y}$ can then be written as
\begin{equation}
\vect{y} = \alpha \vect{x} + \sum_{j=1}^{N_{\rm fg}} \beta_j \vect{f}_j + \sum_{j=N_{\rm fg}+1}^{N_\nu} \gamma_j \vect{g}_j,
\end{equation}
which is to be contrasted with Equation~\eqref{eq:AlphaXBetaZ}.

Again, we can form poor estimators for the amplitudes of the signal ($\hat \alpha = (\vect{x}^T \vect{x})^{-1} \vect{x}^T \vect{\bar y}$), foreground modes ($\vect{\hat \beta} = \mat{F}^T \vect{\bar y}$), and thermal noise modes ($\vect{\hat \gamma}  = \mat{G}^T \vect{\bar y}$). To do so, however, we must first specify how the foreground modes $\mat{F}$ are identified and defined, and in the following section we outline two different methods for this.

\subsection{Two Outlooks on Separation of Signal and Foregrounds}

The first method is closely related to the treatment that we have presented so far. In particular, we considered modes in the matrices $\mat{Z}$ (full rank) or $\mat{F}$ (finite rank) that represent components of the data that are orthogonal to the cosmological signal. There is, however, nothing preventing real foregrounds from being parallel to the signal, and generally $\vect{n}_{\rm fg}$ will be the sum of $ \vect{n}_{{\rm fg}, \parallel \vect{x}} $ (foregrounds parallel to the cosmological signal) and $  \vect{n}_{{\rm fg}, \perp \vect{x}}$ (foregrounds perpendicular to the signal). For example, if the signal and foregrounds share a slowly varying spectral component, it will fall in $\vect{n}_{{\rm fg}, \parallel \vect{x}}$. 

An alternative is to form foreground spectral modes that best describe the covariance of the fluctuating terms on the sky without regard for the cosmological signal. Then, the signal has a piece that is parallel to those foreground modes and a piece that is perpendicular, or $\vect{x} = \vect{x}_{\parallel \mat{F}} + \vect{x}_{\perp \mat{F}}$.

In equations, these two methods are
\begin{eqnarray}
\textrm{ Method 1:}~~~~\vect{y} &=& \vect{x} +  \vect{n}_{{\rm fg}, \parallel \vect{x}} +  \vect{n}_{{\rm fg}, \perp \vect{x}}~~~~\textrm{delete:}~~\vect{n}_{{\rm fg}, \perp \vect{x}} \nonumber \\
\textrm{ Method 2:}~~~~\vect{y} &=& \vect{x}_{\perp \mat{F}} + \vect{x}_{\parallel \mat{F}} + \vect{n}_{\rm fg}~~~~\textrm{delete:}~~\vect{x}_{\parallel \mat{F}} + \vect{n}_{\rm fg}. \nonumber
\end{eqnarray}
In the first method, we can develop a cleaning operation that removes/deweights $\vect{n}_{{\rm fg}, \perp \vect{x}}$, that is, components of the foreground variance orthogonal to the signal. This estimator does not touch the signal by design, so it is guaranteed to have no cosmological signal loss. However, substantial foregrounds will remain in the estimated signal with this rather conservative approach. In the second method, we can develop cleaning that removes any foreground spectral modes that vary spatially on the sky along with any component of the signal parallel to that. If implemented carelessly, this aggressive method will entail cosmological signal loss, although the problem is rectifiable.

Method 1 is described in \citet{rao1967} and is a simply a slight modification of the prescription in Section \ref{sec:cov_from_meas}. Method 2 is the one we develop here and advocate for global $21$\,cm signal recovery.

\begin{figure}
\epsscale{1.2}
\plotone{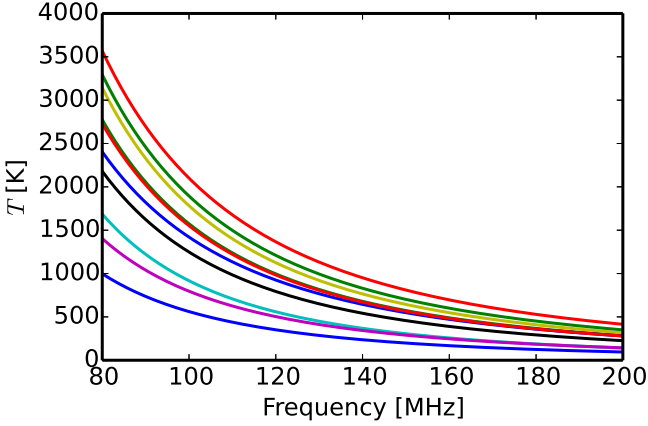}
\caption{Simple simulation of 10 lines of sight with a random combination of three power laws.
\label{fig:simple_sim}}
\end{figure}

We demonstrate the differences in approach using simple simulations. These have $N_\theta$ lines of sight and a number of power laws in the synchrotron emission (to limit the rank). The indices of these power laws are $\{ -2.1, -2.5, -2.9 \}$ and the amplitude is uniformly distributed and positive. Here, $-2.5$ is a characteristic synchrotron index at these frequencies \cite{2012RaSc...47.0K06R, 2014ApJ...782L...9V}, and the two spanning it give additional modes in our initial toy model. Again, as amplitudes we take representative values from the recent observations of \citet{2014ApJ...782L...9V} and scale to the slightly higher frequencies of interest here. For the purposes here we will neglect thermal noise and assume the integration is deep relative to the cosmological signal. Figure~\ref{fig:simple_sim} shows one realization of the foreground model. This model is only meant to be pedagogical, showing the main point of the algorithms here for independent sight lines in a simplified setting with three spectral modes. Beginning in Section~\ref{sec:spatial_information} we will use the full extended GSM model data cubes described in Section~\ref{sec:globsig}.

\begin{figure*}[htb]
\epsscale{1.25}
\includegraphics[scale=0.66]{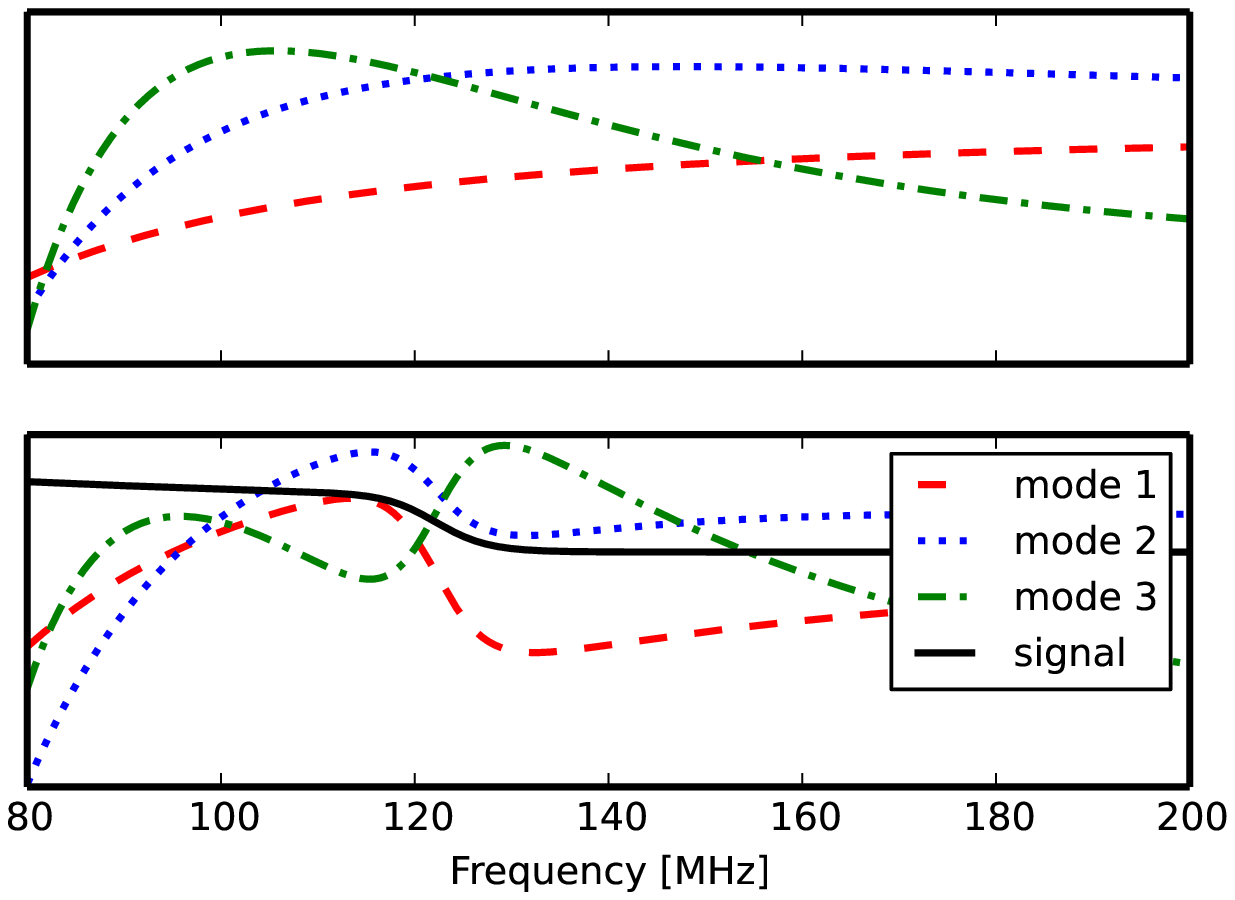}
\includegraphics[scale=0.66]{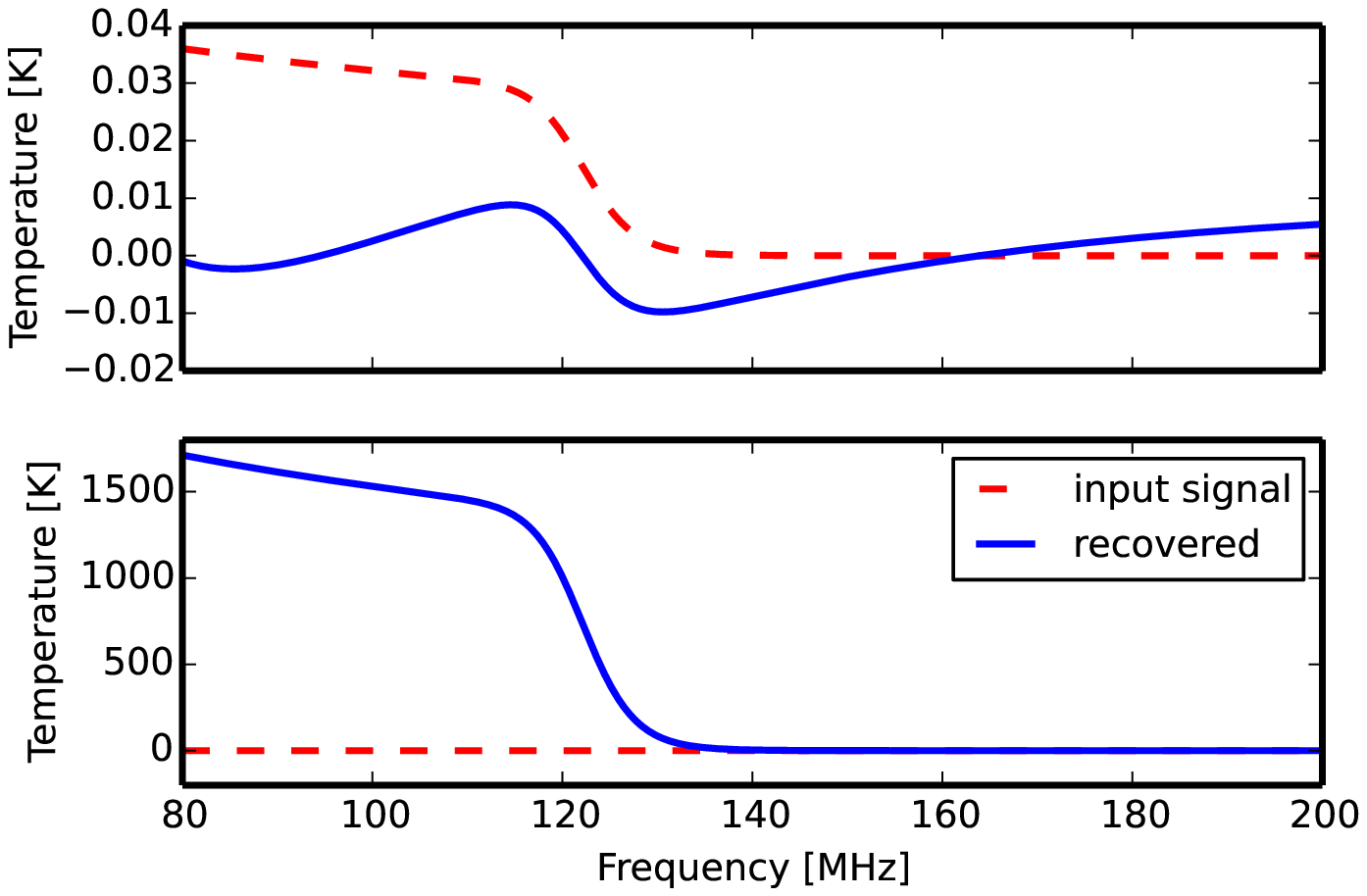}
\caption{Modes discovered from the $(\nu, \nu')$ covariance of simple simulations with three synchrotron spectral indices (Figure~\ref{fig:simple_sim}). These include only the foregrounds and simple signal model of Equation~\eqref{eqn:tanh_model} with $\Delta z = 0.5$ and $z_r = 10.6$. {\it Top left:} the largest three eigenvectors of the covariance $\mathbf{\hat \Sigma}$ that fully describe the foregrounds. {\it Top right:} the filtered data $(1-\mat{F} \mat{F}^T) \vect{\bar y}$. The input signal in red is largely recovered after cleaning the three foreground eigenmodes in the top left. In this toy model, the foregrounds are removed entirely, and only the signal (minus components parallel to the subtracted foregrounds) is left. {\it Bottom left:} the largest three eigenvectors of the restricted covariance $\mathbf{\hat \Sigma} |_{\vect{x}} = [1 - \vect{x} (\vect{x}^T \vect{x})^{-1} \vect{x}^T] \mathbf{\hat \Sigma}$. These are all orthogonal to the cosmological signal. {\it Bottom right:} the data after subtracting the restricted eigenvectors [or $(1-\mat{F}_{\perp} \mat{F}_{\perp}^T) \vect{\bar y}$] are exactly the signal template shape by the construction of the mode functions. The amplitude, however, is clearly dominated by residual foregrounds (note that the signal is too small to be seen on this vertical scale). These are the foreground components parallel to the cosmological signal. While this procedure is immune from loss of cosmological signal, it is insufficiently aggressive when dealing with foregrounds, and the final result is strongly influenced by the foreground covariance. We note that the apparent recovery of the reionization step's location in the bottom-right subplot is spurious, for it is almost entirely driven by our choice of which theoretical template we use.  The location of the ``step" in the top-right subplot, on the other hand, is truly constrained by the data, although signal loss means that it no longer appears as a clean step.
\label{fig:restricted_comparison}}
\end{figure*}

\subsubsection{Method 1: Foreground Modes Orthogonal to Signal}
\label{sss:fore_perp_sig}

To find foreground spectral modes that are orthogonal to the cosmological signal $\vect{x}$, we can find the largest eigenmodes of the restricted covariance
\begin{equation}
\mathbf{\hat \Sigma} \biggl |_{\perp \vect{x}} = [1 - \vect{x} (\vect{x}^T \vect{x})^{-1} \vect{x}^T] \mathbf{\hat \Sigma},
\label{eqn:restricted_cov}
\end{equation}
which in practice will be eigenmodes that represent foregrounds.  By construction, any eigenmode $\vect{f}_i$ of $\mathbf{\hat \Sigma} \bigl |_{\perp \vect{x}} $ will be orthogonal to $\vect{x}$, i.e. $\vect{f}_i^T \vect{x} = 0$. Let $\mat{F}_\perp$ hold the restricted eigenvectors $\vect{f}_i$. The top-left subplot of Figure~\ref{fig:restricted_comparison} shows the unrestricted eigenvectors of an input foreground simulation, while the bottom-left subplot shows the restricted eigenvectors relative to a signal template.

Having identified our foreground eigenvectors, we can construct our estimator for the signal amplitude $\alpha$ using the methods introduced in Section~\ref{sec:cov_from_meas}.  Once again, we first form a poor initial estimator of the signal, which we then adjust using the covariance between the poor estimator and the foregrounds (which is nonzero because a substantial foreground contamination remains in the estimate of the signal).  The covariance of this estimator takes the form
\begin{equation}
{\rm Cov}_{\hat \alpha, \vect{ \hat \beta}} = \left( \begin{array}{cc} (\vect{x}^T \vect{x})^{-1} \vect{x}^T \mathbf{\hat \Sigma} \vect{x} (\vect{x}^T \vect{x})^{-1} & (\vect{x}^T \vect{x})^{-1} \vect{x}^T \mathbf{\hat \Sigma} \mat{F}_\perp \\ \mat{F}^T _\perp\mathbf{\hat \Sigma} \vect{x} (\vect{x}^T \vect{x})^{-1} & \mat{F}^T_\perp \mathbf{\hat \Sigma} \mat{F}_\perp \end{array} \right),
\end{equation}
which looks exactly like Equation~\eqref{eq:ParaPerpCov} but with all occurrences of $\mat{Z}$ replaced by $\mat{F}_\perp$.  This is unsurprising because the modes in $\mat{G}$ represent a diagonal piece of the covariance, $\sigma_{\rm inst}^2 \mat{1}$, and thus are not correlated in frequency with either the foreground modes or the signal.

Projecting out the piece of the signal estimator that is correlated with the foregrounds,
\begin{eqnarray}
\hat \alpha &=& (\vect{x}^T \vect{x})^{-1} \vect{x}^T (1 - \mathbf{\Pi}) \vect{\bar y} \label{eqn:rank-limit_est} \\
\mathbf{\Pi} &=& \mathbf{\hat \Sigma} \mat{F}_\perp (\mat{F}_\perp^T \mathbf{\hat \Sigma} \mat{F}_\perp)^{-1}  \mat{F}_\perp^T.
\end{eqnarray}
That is, project foreground-contaminated modes in the data and then dot against the signal template. The operation $\mathbf{\Pi}$ puts the signal in the basis of foreground modes, weights those by their covariance, and moves back to the spectral space. Note that $\mat{F}_\perp$ does not span the rest of the spectral space like $\mat{Z}$ in the previous section. Here, Equation~\eqref{eqn:rank-limit_est} is not equivalent to $\hat \alpha = (\vect{x}^T \mathbf{\hat \Sigma}^{-1} \vect{x})^{-1} \vect{x}^T \mathbf{\hat \Sigma}^{-1} \vect{\bar y}$.

By construction, our estimator suffers from no loss of cosmological signal (despite the projecting-out of foregrounds) because $(1 - \mathbf{\Pi}) \vect{x} = \vect{x} - \mathbf{\hat \Sigma} \mat{F}_\perp (\mat{F}_\perp^T \mathbf{\hat \Sigma} \mat{F}_\perp)^{-1}  \mat{F}_\perp^T \vect{x} = \vect{x}$.  In the last step, we used the fact that $\mat{F}_\perp^T \vect{x}=0$ through the choice of forming $\mat{F}_\perp$ with eigenvectors restricted to be orthogonal to $\vect{x}$, Equation~\eqref{eqn:restricted_cov}.

The error on $\hat \alpha$ is
\begin{eqnarray}
{\rm Var}(\hat \alpha) &\propto& [\mat{C}_{\parallel, \parallel} - \mat{C}_{\parallel, \perp} \mat{C}_{\perp, \perp}^{-1} \mat{C}_{\perp, \parallel}] \\
&\propto& (\vect{x}^T \vect{x} )^{-1} \vect{x}^T ( 1- \mathbf{\Pi}) \mathbf{\hat \Sigma} \vect{x} (\vect{x}^T \vect{x} )^{-1}.
\label{eqn:alpha_error}
\end{eqnarray}
Here $\vect{x}^T ( 1- \mathbf{\Pi}) \mathbf{\hat \Sigma} \vect{x}$ can be interpreted as projecting all of the foreground modes orthogonal to the signal out of the $(\nu, \nu')$ covariance and finding the noise with respect to that residual covariance. The residual covariance originates from foreground components parallel to the signal and from thermal noise---any $(\nu, \nu')$ covariance component with a nonzero dot product into the signal $\vect{x}$ will contribute to the variance of $\hat \alpha$. This property can be seen in Fig~\ref{fig:restricted_comparison}. The recovered signal has exactly the shape of the signal template $\vect{x}$, but the amplitude is dominated by foregrounds. This feature is clearly undesirable in the regime where foregrounds vastly exceed the signal. 

Another undesirable property of this estimator is that the quoted error ${\rm Var}(\hat \alpha)$ relies on the foregrounds being normally distributed. Real foregrounds are strongly non-Gaussian, and we would like to avoid describing those components in the errors of $\hat \alpha$. We can make a simple modification to the estimator in \citet{rao1967} to treat both of these shortcomings.

\subsubsection{Method 2: Signal Component Orthogonal to Foregrounds}
\label{sss:sig_perp_fore}

In some sense, the method described in Section~\ref{sss:fore_perp_sig} was too conservative.  By limiting our labeling of foregrounds to modes that are orthogonal to the signal, we arrived at an estimator with no formal signal loss, but one in which substantial foreground residuals remained in the final answer.  Here we consider an estimator that more aggressively projects out foregrounds.  The result will be a lossy treatment, but we will also show how this can be rectified.

Instead of isolating the foregrounds orthogonal to the signal, consider the component of the signal that is orthogonal to the foregrounds. The signal is then partitioned as
\begin{equation}
\vect{x} = \vect{x}_{\parallel \mat{F}} + \vect{x}_{\perp \mat{F}}
\end{equation}
For this decomposition to be meaningful, we must once again decide on a definition for our foreground modes.  We let $\mat{F}$ be the largest foreground eigenvectors of $\mathbf{\hat \Sigma}$ (without the orthogonality restrictions of the previous section). In the calculation for $\mathbf{\hat \Sigma}$, the mean $\vect{\bar y}$ contains the cosmological signal and is subtracted out. For this reason, the cosmological signal does not perturb the $(\nu, \nu')$ covariance or its eigenmodes. This is in contrast to inhomogeneous $21$\,cm mapping, where the cosmological signal is stochastic, cannot be subtracted from the covariance, and does perturb the eigenvectors \citep{2013MNRAS.434L..46S, 2013ApJ...763L..20M}. (The foreground modes can optionally be isolated from thermal noise by forming the $(\nu, \nu')$ cross-variance of maps acquired at two different times, as was done in \citealt{2013MNRAS.434L..46S}.)

To project the identified foreground modes out of the estimator, we apply $1 - \mathbf{\Pi} = 1 - \mat{F} \mat{F}^T$.  This operation splits the spectrum into a foreground-contaminated and a (theoretically) foreground-clean subspace as
\begin{equation}
\left( \begin{array}{c} \mathbf{\Pi} \\ 1- \mathbf{\Pi} \end{array} \right) \vect{y} = \left( \begin{array}{c} \mathbf{\Pi} \\ 1 - \mathbf{\Pi} \end{array} \right) \alpha \vect{x} +
\left( \begin{array}{c} \vect{n}_{\rm fg} \\ \mathbf{0} \end{array} \right) + \vect{n}_{\rm inst}.
\end{equation}

Unlike the previous estimator Equation~\eqref{eqn:rank-limit_est}, however, the foreground modes are no longer constructed to be orthogonal to the signal.  By projecting out the foregrounds, then, it is likely that some cosmological signal will be lost.  Mathematically, $(1 - \mat{F} \mat{F}^T) \vect{x}$ equals $\vect{x}_{\perp \mat{F}}$, not $\vect{x}$, and the estimator $\hat \alpha = (\vect{x}^T \vect{x})^{-1} \vect{x}^T (1 - \mathbf{\Pi}) \vect{\bar y}$ is an incorrectly normalized estimator of the signal amplitude (i.e., it suffers from a multiplicative bias). Correcting this, we propose the revised estimator
\begin{equation}
\hat \alpha = \frac{\vect{x}^T (1 - \mat{F} \mat{F}^T) \vect{\bar y}}{\vect{x}^T (1 - \mat{F} \mat{F}^T) \vect{x}}.
\label{eqn:iid_limited_est}
\end{equation}

In the error calculation from Equation~\eqref{eqn:alpha_error}, the key term $( 1- \mathbf{\Pi}) \mathbf{\hat \Sigma} $ now reduces to $ \mathbf{1} \sigma_{\rm inst}^2$, under the assumption that all foregrounds are removed to a good approximation and only Gaussian thermal noise remains. While the foregrounds may be strongly non-Gaussian, we will assume that the residuals after the modes $\mat{F}$ have been subtracted are Gaussian and due to the thermal noise floor. This would need to be verified in an analysis of real data. 

The overall error is now
\begin{equation}
{\rm Var}(\hat \alpha) \sim \left ( 1 + \frac{N_{\rm fg}}{N_\theta - 1 - N_{\rm fg}} \right ) \frac{\sigma_{\rm inst}^2}{[\vect{x}^T (1 - \mat{F} \mat{F}^T) \vect{x}]^2}
\label{eqn:gaussian_mode_error}
\end{equation}

To get some rough intuition, assume that the foreground modes all have about the same overlap with the cosmological signal. Then $\vect{x}^T (1 - \mat{F} \mat{F}^T) \vect{x} \propto N_\nu - N_{\rm fg}$. As $N_{\rm fg}$ approaches $N_\nu$, the foreground modes do a progressively better job of spanning the $N_\nu$ spectral space. In the limit that they fully span the space (there is no signal distinguishable from foregrounds), the signal is also nulled and the error diverges. Alternately, as $N_{\rm fg}$ approaches $N_\theta$, the foreground discovery process uses up all of the spatial degrees of freedom and the error diverges. In this formulation, then, we are self-consistently including the possibility of signal loss in our error analysis.

\section{Constraints on Passband Calibration}
\label{sec:instrum}

The abstract design guidance suggested by this method is that (1) the experiment should not increase the rank of the foreground covariance (keep $N_{\rm fg}$ small) and (2) if there are foregrounds with $N_{\rm fg}$ spectral degrees of freedom, at least $N_\theta = N_{\rm fg}$ samples of its variations need to be observed if those foregrounds are to be subtracted.  

A time-varying passband calibration is the worst instrumental systematic in this sense. Here each line of sight sees a slightly different foreground spectrum and so requires a different spectral function for cleaning. In the limit that each line of sight can have a different foreground,  $N_{\rm fg} \rightarrow N_\theta$ and the errors diverge.  In contrast, a constant passband error only modulates the foreground covariance spectral functions, but it does not increase the rank of its covariance. Its impact is more aesthetic, can be corrected by recalibrating off the brightest mode, and does not fundamentally limit the investigation. (This is proven in Section~\ref{sec:separable_cov} using the specific notation developed there.)

\begin{figure*}[htb]
\epsscale{1.25}
\includegraphics[scale=0.6]{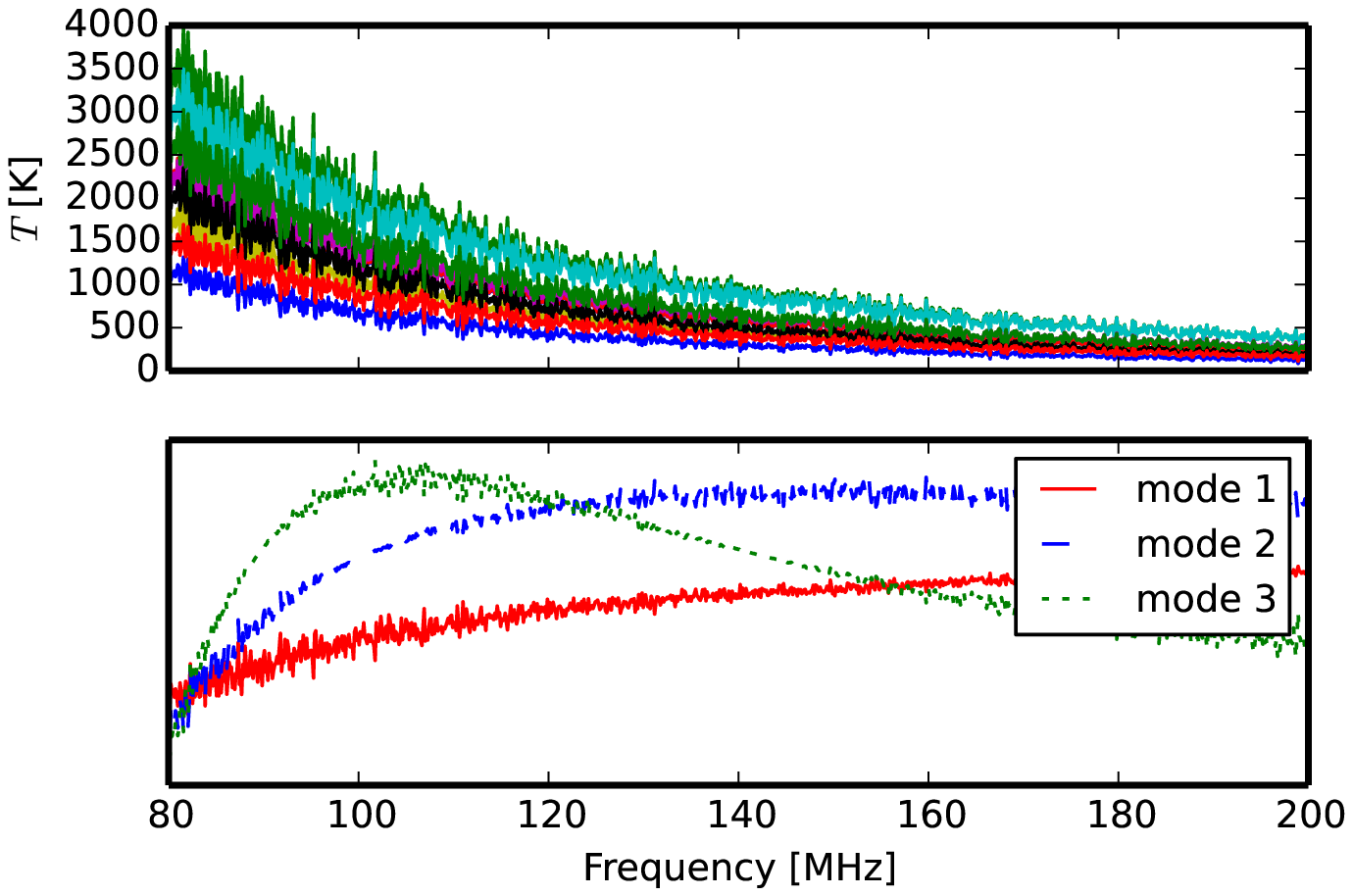}
\includegraphics[scale=0.6]{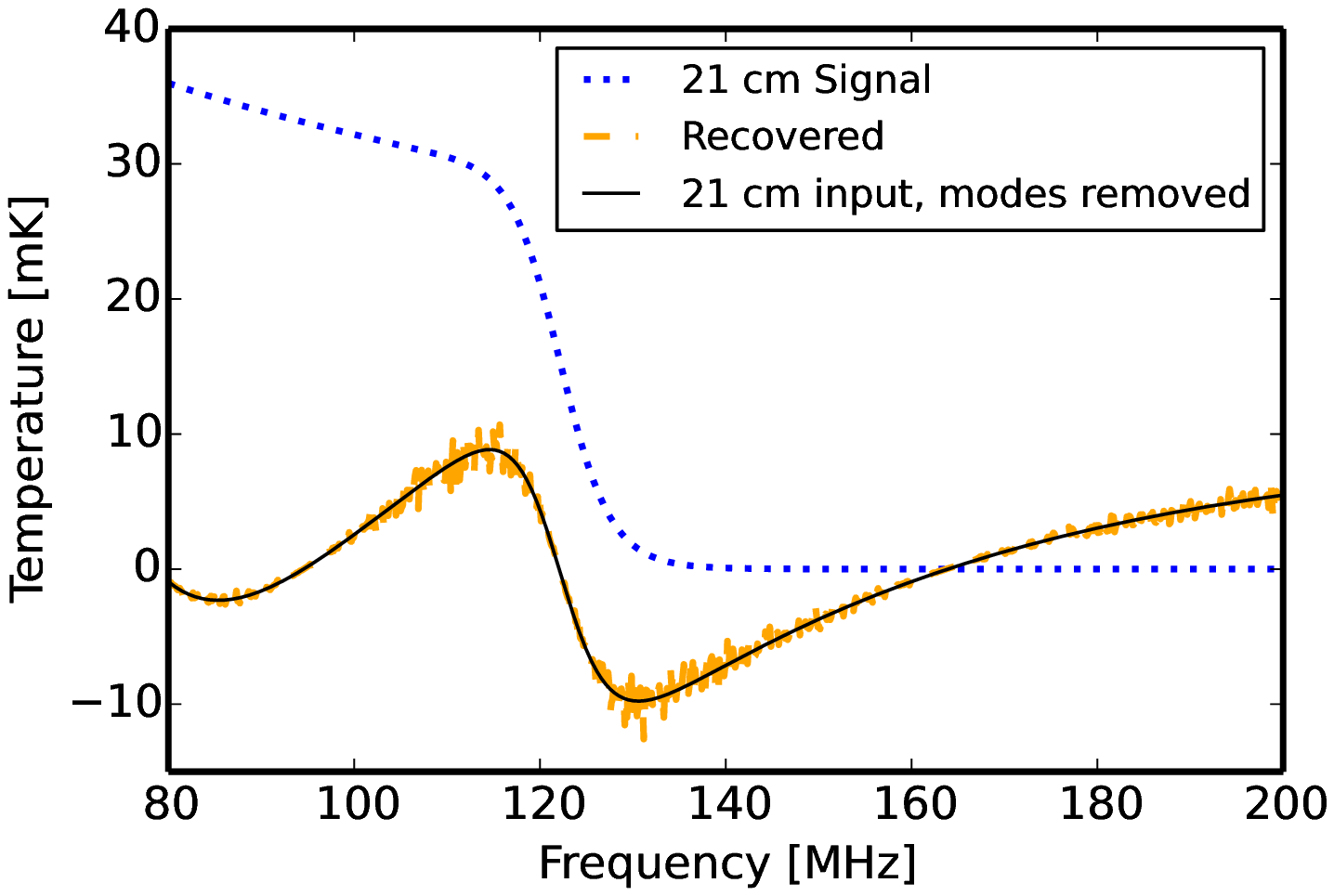}
\caption{Toy model showing that an absolute passband calibration error that does not vary with time does not impact primordial signal recovery. {\it Top left:} observations of the spectrum across 10 lines of sight with a stable $10\%$ passband calibration error between bins (here neglecting thermal noise). The input synchrotron is taken from a mixture of three power laws, so it is described by three spectral modes in the data. {\it Bottom left:} foreground modes discovered in the data. Because of passband calibration error, these modes are not smooth. A constant calibration error does not increase the rank of the three input power laws. {\it Right:} recovered cosmological signal compared to the input signal, and the cosmological signal with the three modes on the left removed. The fixed calibration error is cosmetic.
\label{fig:fixed_cal_error}}
\end{figure*}

Figure~\ref{fig:fixed_cal_error} shows the 10 simulated lines of sight with a constant $10\%$ calibration error. Here we multiply the spectrum along each line of sight by a constant $1+\vect{\delta}$, where $\vect{\delta}$ is $N_\nu$ long and normally distributed. Fitting a smooth power law to data with this constant $1+\vect{\delta}$ calibration error would leave unacceptable residuals, but the modes are able to discover the constant calibration error and subtract the foreground with modest signal loss. This parameterization of a calibration error is a matter of convenience. In simulations where the errors $\vect{\delta}$ are a smoothly varying function of frequency or even have the same shape as the signal template, the conclusions are the same. (In the pathological case that the calibration error interacts with the foreground power law to produce modes identical to the cosmological template, indeed signal would be lost. This loss is self-consistently accounted for in the method and could be discovered by visual inspection of the modes.)
 
\begin{figure}
\epsscale{1.2}
\plotone{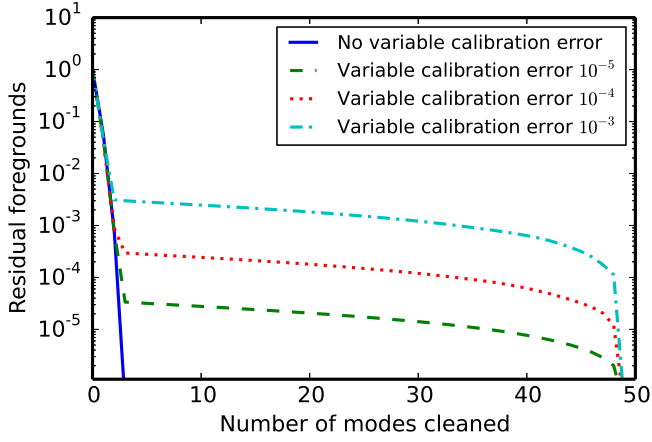}
\caption{Residual foregrounds as a function of modes removed in a toy model with three foreground degrees of freedom. Amplitudes are normalized to one for zero modes removed and in rms rather than variance. In the case with no variable calibration error, the variance in the spectrum quickly reaches thermal noise after three modes are removed. If the passband calibration varies between lines of sight, the residuals become significantly worse. Generally, for foregrounds $10^5$ times larger than the signal, the passband should be stable to $\sim 10^{-5}$. Passband calibration instability spreads the variance of the three intrinsic foreground modes across many more degrees of freedom. 
\label{fig:variable_cal_eigenvalues}}
\end{figure}

In contrast, Figure~\ref{fig:variable_cal_eigenvalues} shows that even a $0.1\%$ \emph{variable} passband calibration error spreads foreground variance into many modes that are all significantly larger than the signal ($10^{-5}$ there). To implement a variable passband, we multiply each line of sight by a different $1+\vect{\delta}_i$, representing a change in the instrument response with time. 

The rule of thumb here is that if foregrounds are $R$ times larger than the signal, then the per-pointing calibration needs to be stable to $\sim 1/R$. Averaging over sight lines tends to relax this constraint. In the simulations shown in Section~\ref{sec:res_elt}, we find that the stability must be better than $\sim 10^{-4}$ to have no effect. The constraint degrades rapidly above that. In the other limit of extremely high stability, we find that once calibration stability is suppressed below the thermal noise, there is no further gain in foreground discrimination.

An important point is that the method developed here can discover any response to foregrounds that varies across the sky (non-monopole). This does not necessarily need to originate from passband calibration or differential polarization response---these are just plausible sources of error from instrumental response. 

\section{Incorporating Spatial Information}
\label{sec:spatial_information}

The previous discussion assumed that the foregrounds are stationary and independent between different lines of sight.  It is for this reason that the first step in the prescription outlined above was to form $\overline{y}$, the unweighted average of all lines of sight---with stationary statistics, there is no reason to prefer one direction of the sky over another, and an unweighted average provides the best signal-to-noise ratio.  This is the implicit assumption that is being made by experiments that average over large regions of sky at once (such as single-dipole experiments) and produce a single spectrum as their measurement.  The results in preceding sections therefore also apply to experiments with no angular sensitivity, except without that angular information, error bars and foreground cleaning methods cannot be informed by the data itself and must be derived a priori.

In reality, foregrounds are neither stationary nor independent.  The synchrotron brightness varies across the sky (violating stationarity), and in addition the foregrounds are known to be spatially correlated (violating independence).

Angular information can be leveraged in several ways.  First, nonstationarity of the foregrounds allows the estimator to down-weight parts of the sky that are known to be particularly noisy, such as the galactic center.  Second, correlation information allows foreground properties in one part of the sky to be inferred from observations of another part of the sky.  For example, in the unrealistically extreme limit of perfectly correlated foregrounds, a measurement of foregrounds in any part of the sky automatically allows a perfect prediction of foreground brightness in any other part of the sky.

To take advantage of angular information, L13 considered an optimal estimator given full-sky maps at all frequencies of interest as well as a known $\nfreq \npix \times \nfreq \npix$ covariance $\mathbf{N}$ between all pixels and all frequencies.  Let $\vect{d}$ be a vector of length $N_\nu N_\theta$ containing the measured sky maps.  Here we use $\npix$ to denote the number of pixel indices in a full-sky HEALPix map \citep{2005ApJ...622..759G}. For reasons we will see in Section~\ref{sec:res_elt}, this also prevents confusion between the $N_\theta$ independent sight lines in the previous section and the present discussion, which is made more complex by beam convolution.

The observations $\vect{d}$ are related to the global $21$\,cm spectrum $\vect{s}$ via the measurement equation
\begin{equation}
\label{eq:PawnToE4}
\vect{d} = \mat{A} \vect{s} + \vect{n},
\end{equation}
where $\vect{n} \equiv \nfg + \ninst$ is the generalized noise, containing $\nfg$ and $\ninst$ as the foreground and instrumental noise contribution to $\vect{d}$, respectively.  The $\npix \nfreq \times \nfreq$ matrix $\mat{A}$ is given by $\mat{1} \otimes \vect{e}_0$, where $\otimes$ is the Kronecker product, $\mat{1}$ is an $\nfreq \times \nfreq$ identity matrix, and $\vect{e}_0$ is the spatial monopole, i.e., an $\npix$-long vector of 1\,s.  Its function is to copy, at every frequency, the value of the global spectrum $\vect{s}$ to all of the spatial pixels comprising the sky map.  The maximum-likelihood estimate $\hat{\mathbf{s}}_\textrm{ML}$ for the full global spectrum is given by
\begin{equation}
\label{eq:UnrealisticOptEst}
\hat{\vect{s}}_\textrm{ML} = (\mat{A}^T \mat{N}^{-1} \mat{A})^{-1} \mat{A}^T \mat{N}^{-1} \vect{d},
\end{equation}
where $\mat{N} \equiv \langle \vect{n} \vect{n}^T \rangle - \langle \vect{n} \rangle \langle \vect{n} \rangle^T$ is the covariance matrix of the generalized noise, which is assumed to be known.  To facilitate comparisons with previous work, we will begin by considering the estimator $\hat{\vect{s}}$ of the full global spectrum. Section~\ref{sec:separable_cov} returns to the template amplitude constraint.

\subsection{Spatial Weighting to Recover the Spectrum}

L13 used a series of analytical manipulations to show that incorporating angular information via Equation~\eqref{eq:UnrealisticOptEst} could in principle lead to large reductions in foreground contamination.  In practice, however, this prescription is difficult to implement for a number of reasons.  First, one may not possess sufficiently accurate models of the instrument and the foregrounds to write down the matrix $\mat{N}$, placing a priori expectations on the instrument response. This matrix cannot be derived from the data (unlike $\boldsymbol \Sigma$ from previous sections) because it has dimensions $N_\nu N_\theta \times N_\nu N_\theta$ and therefore contains many more degrees of freedom than the number of measurements $N_\nu N_\theta$.  Moreover, even if the matrix is somehow known, its large size makes its inversion in Equation~\eqref{eq:UnrealisticOptEst} computationally challenging.

To deal with these challenges, consider a modified recipe where the sky maps are dealt with frequency by frequency (unlike in Equation~\eqref{eq:UnrealisticOptEst} where all frequencies and all lines of sight are mixed together) and an estimator for the global signal at each frequency is formed by spatially averaging with nontrivial weights (unlike in Equation~\eqref{eq:StraightAverage}, where different lines of sight were equally weighted).  Let $\vect{d}_\beta$ be a vector of length $\npix$ that represents the map within $\vect{d}$ corresponding to the $\beta$th frequency channel.  The global signal at this frequency channel can then be estimated by computing the weighted average
\begin{equation}
\label{eq:DotProd}
\hat{\vect{s}}_\beta = \vect{w}_\beta^T \vect{d}_\beta,
\end{equation}
where $\vect{w}_\beta$ is a vector of length $\npix$ with weight appropriate for the $\beta$th frequency channel. Throughout this paper, we adopt a convention where summations are written explicitly. Repeated indices therefore do \emph{not} imply summations. The variance of this estimator---which we will seek to minimize---is
\begin{equation}
\label{eq:SepEstVar}
\textrm{Var}(\hat{\vect{s}}_\beta) \equiv \langle \hat{\vect{s}}_\beta^2 \rangle - \langle \hat{\vect{s}}_\beta \rangle^2 = \vect{w}_\beta^T \mathbf{\Phi}_\beta \vect{w}_\beta,
\end{equation}
where $\mathbf{\Phi}_\beta$ is the $\npix \times \npix$ \emph{spatial} covariance matrix of the map at the $\beta$th frequency.  For our final estimator to be correctly normalized, we require the weights at each frequency to sum to unity, i.e., $\vect{w}_\beta^T \vect{e}_{0} = 1$.  To obtain the minimum-variance solution for our weights, we use a Lagrange multiplier $\mu$ to impose our normalization constraint and minimize the quantity
\begin{equation}
\label{eq:FirstLagrange}
L = \vect{w}_\beta^T \mathbf{\Phi}_\beta \vect{w}_\beta - \mu \vect{w}_\beta^T \vect{e}_{0}.
\end{equation}
Doing so gives
\begin{equation}
\label{eq:SpatialWeights}
\vect{w}_\beta = \frac{\mathbf{\Phi}^{-1}_\beta \vect{e}_{0}}{\vect{e}_{0}^T \mathbf{\Phi}^{-1}_\beta \vect{e}_{0}},
\end{equation}
and therefore
\begin{equation}
\label{eq:weightedspectrum}
\hat{\vect{s}}_\beta = \frac{\vect{e}_{0}^T \mathbf{\Phi}^{-1}_\beta \vect{d}_\beta}{\vect{e}_{0}^T \mathbf{\Phi}^{-1}_\beta \vect{e}_{0}}.
\end{equation}
In words, these weights tell us to whiten the sky maps at each frequency using the inverse spatial covariance before summing together different lines of sight and normalizing. Because the whitening takes the form of a nondiagonal matrix multiplication, this operation not only down-weights brighter parts of the sky but also makes use of angular correlation information to better estimate the monopole signal. \citet{2003PhRvD..68l3523T} tackled the transpose of this problem for CMB foreground removal. Note that in modeling $\boldsymbol \Phi_\beta$, it is essential to capture the nonstationary nature of our galaxy's foreground emission.  It is insufficient, for example, to describe the galaxy using an angular power spectrum alone, which assumes statistical isotropy. This would not only be physically unrealistic but would also make our weights constant across the sky (a fact that can be derived by applying Parseval's theorem to our expression for $\vect{w}_\beta$), defeating the purpose of using spatial weights in the first place.

\subsection{Spatial Weights Within a Separable Covariance Model}
\label{ssec:separable_intro}

The spatial weight recipe that we have just specified can also be considered a special limit of the maximum-likelihood estimator, albeit with one small modification. Suppose that the full $N_\nu N_\theta \times N_\nu N_\theta$ covariance $\mat{N}$ is separable so that we can write it as $\boldsymbol \Sigma \otimes \boldsymbol \Phi$, where $\boldsymbol \Sigma$ is the $\nfreq \times \nfreq$ spectral covariance from previous sections, $\boldsymbol \Phi$ is an $\npix \times \npix$ spatial covariance, and $\otimes$ is the Kronecker product.

Using the identities $(\mat{G} \otimes \mat{H} )^{-1} \equiv \mat{G}^{-1} \otimes \mat{H}^{-1}$ and $(\mat{G} \otimes \mat{H})(\mat{J} \otimes \mat{K}) \equiv \mat{G}\mat{J} \otimes \mat{H}\mat{K}$, we have
\begin{equation}
\label{eq:AtNinv}
\mat{A}^T \mat{N}^{-1} = \boldsymbol \Sigma^{-1} \otimes \vect{e}_0^T \boldsymbol \Phi^{-1}
\end{equation}
and
\begin{equation}
\label{eq:AtNinvAinv}
(\mat{A}^T \mat{N}^{-1} \mat{A})^{-1} = \boldsymbol \Sigma \otimes (\vect{e}_0^T \boldsymbol \Phi^{-1} \vect{e}_0)^{-1}.
\end{equation}
Inserting these into Equation~\eqref{eq:UnrealisticOptEst}, the final estimator in this separable approximation is then
\begin{equation}
\hat{\vect{s}}_{\beta} \Bigg{|}_\textrm{sep} = \frac{\vect{e}_0^T \boldsymbol \Phi^{-1} \vect{d}_\beta}{\vect{e}_0^T \boldsymbol \Phi^{-1} \vect{e}_0},
\end{equation}
which is almost identical to Equation~\eqref{eq:weightedspectrum}.  If we slightly relax the assumption of perfect separability by allowing $\boldsymbol \Phi$ to acquire a frequency dependence (so that $\boldsymbol \Phi \rightarrow \boldsymbol \Phi_\beta$), the correspondence becomes exact. Including this frequency-dependent nonseparability also allows us to incorporate the fact that the instrumental beam will generally broaden toward lower frequencies, in addition to intrinsic nonseparability of the foregrounds. Section~\ref{ssec:block_separable} discusses nonseparability caused by the differences between foregrounds and thermal noise.

Interestingly---but unsurprisingly---the spectral covariance $\boldsymbol \Sigma$ drops out of the optimal estimator for the global spectrum once separability is invoked.  As discussed in L13, this is due to the fact that once we spatially average over the maps, we are left with a spectrum consisting of $\nfreq$ numbers.  If our final goal is to measure a cosmological spectrum, that is also of length $\nfreq$, the constraint that our estimator be correctly normalized and unbiased means that there is nothing left to do.  Mathematically, this manifested itself in our derivation when the factors of $\boldsymbol \Sigma$ in Equations~\eqref{eq:AtNinv} and \eqref{eq:AtNinvAinv} canceled each other out when forming $(\mat{A}^T \mat{N}^{-1} \mat{A})^{-1} \mat{A}^T \mat{N}^{-1}$.

If we return to our method in previous sections, however, and attempt to constrain the amplitude $\alpha$ of a theoretical template, the spectral covariance $\boldsymbol \Sigma$ re-enters the discussion.  When constraining $\alpha$, one is seeking to measure a single number from $\nfreq$ measurements, which in general ought to be combined nonuniformly if these measurements have different error properties (as captured by $\boldsymbol \Sigma$). 

\section{Analyzing Global $21$\,cm Data Cubes}
\label{sec:separable_cov}

In this section we combine the finite-rank, in situ estimate of spectral foregrounds with a model of angular correlations. The end product will be an estimator for the amplitude of a cosmological signal template that can be applied to experimental global $21$\,cm data cubes. The angular part of the estimator uses assumed spatial distributions to improve the amplitude estimate. In contrast, the central aspect of this estimator is that it does not assume a set of \emph{spectral} modes or spectral covariance of the data.

It is notationally convenient to write the $\npix \nfreq$-long data matrix $\vect{d}$ by reshaping it into an $\nfreq \times \npix$ matrix $\vect{Y}$. This ``data cube'' is a stack of maps at all of the observed frequencies. The original data vector $\vect{d}$ can be recovered through the serializing ``vec()'' operation ${\rm vec}(\mat{Y}) = \vect{d}$. This data cube form naturally accommodates weighting operations by the separable covariance. A left-multiplication acts on the spectral direction, and a right-multiplication acts on the spatial direction.

The essential weighting operation on the full data set is $\mat{N}^{-1} \vect{d}$, which under the separable covariance assumption becomes
\begin{equation}
\mat{N}^{-1} \vect{d} = (\mathbf{\Sigma} \otimes \mathbf{\Phi})^{-1} \vect{y}  = \mathbf{\Sigma}^{-1} \mat{Y} \mathbf{\Phi}^{-1}
\end{equation}
using identities of the Kronecker and vec operations.

A model for the noise on $\mat{Y}$ is to draw some matrix $\mat{E}$ with shape $N_\nu \times N_{\rm pix}$ of normal deviates from $N(\mu=0,\sigma^2=1)$ and then correlate the spectral and spatial parts as $\mathbf{\Sigma}^{1/2} \mat{E} \mathbf{\Phi}^{1/2}$ to give a noise realization drawn from the covariance $\mathbf{\Sigma} \otimes \mathbf{\Phi}$.

The data model in this presentation is 
\begin{equation}
\mat{Y} = \alpha \vect{x} \vect{e}_0^T + \mathbf{\Sigma}^{1/2} \mat{E} \mathbf{\Phi}^{1/2}.
\end{equation}
This is a specialized case of the ``growth curve'' model \citep{KolloRosen}, reviewed in Appendix~\ref{sec:growth_curve}. We can transfer the separable method developed in Section~\ref{ssec:separable_intro} to the present case of constraining the template amplitude $\alpha$ by letting $\vect{s} \rightarrow \alpha$ and $\mat{A} = \mat{1} \otimes \vect{e}_0 \rightarrow \vect{x} \otimes \vect{e}_0$ (because we are now assuming a form for $\vect{x}$ as a theoretical template). Here, the maximum-likelihood estimate for the amplitude of some spectral template $\vect{x}$ is 
\begin{equation}
\hat \alpha = (\vect{x}^T \mathbf{\Sigma}^{-1} \vect{x})^T \vect{x}^T \mathbf{\Sigma}^{-1} \mat{Y} \mathbf{\Phi}^{-1} \vect{e}_0 (\vect{e}_0^T \mathbf{\Phi}^{-1} \vect{e}_0)^{-1}.
\end{equation}
This applies the familiar maximum-likelihood estimators for the spatial monopole (with respect to covariance $\mathbf{\Phi}$) and spectral template (with respect to covariance $\mathbf{\Sigma}$) to the right and left side of our data cube $\mat{Y}$. 

Following Section~\ref{sec:cov_from_meas}, we can seek to replace $\mathbf{\Sigma}$ with a $(\nu, \nu')$ covariance measured in the data, $\mathbf{\hat \Sigma}$. In Section~\ref{sec:cov_from_meas}, there was no information distinguishing different lines of sight, so the mean subtraction in Equation~\eqref{eqn:sample_cov} simply used the unweighted average of $\vect{y}_i$. We now have information about the angular covariance and can improve the mean spectrum estimation by including the spatial weight, as $\mat{Y}_{\rm sub}^T = (1 - \vect{e}_{0} \vect{w}^T) \mat{Y}^T$. Here $\vect{w}= \mat{\Phi}^{-1} \vect{e}_{0} (\vect{e}_{0}^T \mat{\Phi}^{-1} \vect{e}_{0})^{-1}$ is an $N_{\rm pix}$-long weight map. Another way of looking at the operation $(1 - \vect{e}_{0} \vect{w}^T)$ is that it subtracts the monopole spectrum out of the map so that the eigenvectors of $\mathbf{\hat \Sigma}$ will only contain spatially variable spectral modes, not the monopole. Procedurally, one (1) applies the angular weight $\mat{\Phi}^{-1}$, (2) dots against the spatial monopole template $\vect{e}_{0}$, (3) accounts for the impact of the spatial weight through $(\vect{e}_{0}^T \mat{\Phi}^{-1} \vect{e}_{0})^{-1}$, and 4) projects this spectral template as a monopole across the map (under $\vect{e}_{0}$) and subtracts this from the measured map. Using the mean-subtracted map, we can find the weighted covariance as
\begin{equation}
\mathbf{\hat \Sigma} = \mat{Y}_{\rm sub} \mat{\Phi}^{-1} \mat{Y}_{\rm sub}^T~~{\rm with}~~\mat{Y}_{\rm sub}^T = (1 - \vect{e}_{0} \vect{w}^T) \mat{Y}^T.
\end{equation}

The estimator for the monopole spectrum template amplitude is then
\begin{equation}
\hat \alpha = (\vect{x}^T \mathbf{\hat \Sigma}^{-1} \vect{x})^{-1} \vect{x}^T \mathbf{\hat \Sigma}^{-1} \mat{Y} \vect{w}
\end{equation}
\citet{KolloRosen} show that this choice of weight yields the maximum-likelihood $\hat \alpha$. The inverse $\mathbf{\Phi}^{-1}$ of the $N_{\rm pix} \times N_{\rm pix}$ spatial covariance appears explicitly only in the construction of the sample variance. Elsewhere, it is collapsed onto the single simple weight map $\vect{w}$. Section~\ref{ssec:angular_weighting} describes choices for the form of $\mat{\Phi}$. There we advocate a diagonal form of $\mat{\Phi}$ based on thermal noise, whose amplitude is determined by a position-dependent sky temperature and integration time. Diagonal $\mat{\Phi}$ makes estimation of $\mathbf{\hat \Sigma}^{-1}$ computationally simple.
%One approximation that can be made in practice is to replace $\mathbf{\Phi}^{-1}$ with its diagonal, which simply weighs angles differently in the $(\nu, \nu')$ variance estimation. 

We will argue in Section~\ref{ssec:angular_weighting} that the angular weights only need to treat the residual foregrounds and thermal noise variations across the sky. We choose to let $\mathbf{\hat \Sigma}^{-1}$ do the legwork of determining the spectral modes to remove from the data itself. In this role, $\mathbf{\hat \Sigma}^{-1}$ accounts for instrumental systematics, whereas $\mathbf{\Phi}^{-1}$ can improve the estimate of $\hat \alpha$ by using a foreground model to down-weight noisier regions.

Similar to the discussion in Section~\ref{sec:limited_rank_iid}, we can replace the $\mathbf{\hat \Sigma}^{-1}$ operation by a projection that removes a finite number of contaminated spectral modes. This is equivalent to (1) identifying a data subspace parallel to foregrounds that is contaminated and fully removed and (2) keeping a data subspace orthogonal to foregrounds. 

Let $\mat{F}$ contain the largest eigenvectors of the sample covariance $\mathbf{\hat \Sigma}$. Then
\begin{equation}
\hat \alpha = \frac{\vect{x}^T (1- \mat{F} \mat{F}^T) }{\vect{x}^T (1- \mat{F} \mat{F}^T) \vect{x}} \mat{Y} \vect{w}.
\label{eqn:est_ang_freq}
\end{equation}
The denominator $\vect{x}^T (1- \mat{F} \mat{F}^T) \vect{x}$ is a scalar normalization.

This is analogous to Equation~\eqref{eqn:iid_limited_est}, except that instead of a simple $\vect{\bar y}$, each spatial slice of the map $\mat{Y}$ is weighted to find the mean as $\mat{Y} \vect{w}$.  

Using this notation, we can easily prove that constant passband calibration errors do not increase the rank of foreground spectral modes. Let $1+\delta \vect{c}$ be an $N_\nu$-long vector that represents constant miscalibration of the passband. If a constant calibration error multiplies all lines of sight, $\mat{C} = {\rm diag}(1+\delta \vect{c})$ modifies the data as $\mat{C} \mat{Y}$. So long as the response is not zeroed out at some frequencies, $\mat{C}$ is invertible, and ${\rm rank}(\mat{C} \mathbf{\hat \Sigma} \mat{C}^T) = {\rm rank}(\mathbf{\hat \Sigma})$ for invertible $\mat{C}$. Intuitively, a constant passband error can be thought of as just a per-frequency rescaling of units. Because such a rescaling does not affect the rank of the foregrounds, there is no increase in the number of foreground modes that need to be constrained. Our self-discovery scheme for spectral modes will therefore be just as effective when constant passband miscalibrations are present.

\subsection{Deriving the Final Estimator: A Two-component Covariance}
\label{ssec:block_separable}

In deriving Equation~\eqref{eqn:est_ang_freq}, the key assumption that we made was that the total covariance is separable into spatial and spectral components. This confuses foregrounds and thermal noise, which each have very different angular distributions. For example, we know that a synchrotron spectral mode is associated with broad spatial distributions of galactic emission. In contrast, thermal noise is always uncorrelated across angular pixels, but may change in variance across the sky. In the limit that spectral foreground subtraction is perfect, only Gaussian thermal noise will remain. In contrast, the separable assumption would dictate that it has approximately the same angular correlations as the galaxy. This issue ultimately translates into the angular weight $\vect{w}$ that should be applied. 

The solution to this problem is to let the total covariance be nonseparable but comprised of independently separable foreground and noise components. Here, $\mat{N} = \mat{N}_f + \mat{N}_g$ where $\mat{N}_f = \mathbf{\Sigma}_f \otimes \mathbf{\Phi}_f$ and $\mat{N}_g = \mathbf{\Sigma}_g \otimes \mathbf{\Phi}_g$.  For concreteness, imagine that the ``$f$" component consists of modeled foreground modes in $\mat{F}$, and the ``$g$" component consists of any residuals after the projection operation. This is exactly analogous to the $\mat{F}/\mat{G}$ split in \citet{rao1967} described in Section~\ref{sec:limited_rank_iid}. If our discovered foreground modes happen to perfectly capture the true foregrounds, then the ``g'' component would consist only of Gaussian thermal noise. In general, however, we need not make this assumption.

The central problem with the model $\mat{N} = \mathbf{\Sigma} \otimes \mathbf{\Phi}$ is that $(1-\mat{F} \mat{F}^T)$ only acts on the $\mathbf{\Sigma}$ part but not on $\mathbf{\Phi}$. In contrast, in the two-component model, when $(1-\mat{F} \mat{F}^T)$ acts on the data, the remaining covariance is only $\mathbf{\Sigma}_g \otimes \mathbf{\Phi}_g$, e.g., it correctly ascribes $\mathbf{\Phi}_f$'s spatial correlations to the modes that $\mat{F}$ removes. 

We can now reassess the $\vect{w}$ in Equation~\eqref{eqn:est_ang_freq} that yields the minimum-variance estimate of $\alpha$. To clarify the discussion, let $\vect{q} \equiv [\vect{x}^T (1- \mat{F} \mat{F}^T) \vect{x}]^{-1}  (1- \mat{F} \mat{F}^T) \vect{x}$.  The first step of Equation~\eqref{eqn:est_ang_freq} in this notation is to form $\vect{q}^T \vect{Y}$, which can be interpreted as a series of estimators for $\alpha$, one for each line of sight.

With our two-component covariance, a little algebra reveals that the variance to minimize is given by
\begin{equation}
\textrm{Var}(\hat{\alpha}) = (\vect{q}^T \boldsymbol \Sigma_f \vect{q})(\vect{w}^T \boldsymbol \Phi_f \vect{w}) + (\vect{q}^T \boldsymbol \Sigma_g \vect{q})(\vect{w}^T \boldsymbol \Phi_g \vect{w}).
\end{equation}
The first term is zero by construction because $\vect{q}^T \boldsymbol \Sigma_f \propto (\mat{1} - \mat{F}\mat{F}^T) \mat{F} =0$.  Minimizing the remaining term subject to the constraint that the weights sum to unity requires minimizing the quantity
\begin{equation}
L = (\vect{q}^T \boldsymbol \Sigma_g \vect{q})(\vect{w}^T \boldsymbol \Phi_g \vect{w}) - \mu  \vect{w}^T \vect{e}_{0},
\end{equation}
where $\mu$ is again a Lagrange multiplier. Because $\vect{q}^T \boldsymbol \Sigma_g \vect{q}$ is just a constant, it can be absorbed into $\mu$ by a simple rescaling.  One then sees that $L$ is identical to Equation~\eqref{eq:FirstLagrange}, except that $\boldsymbol \Phi_g$ (the spatial covariances of the residuals) takes the place of $\boldsymbol \Phi_\beta$ (the full spatial covariance at the $\beta$th frequency).  We may thus import our previous solution for the spatial weights to find
\begin{equation}
\label{eq:FinalWeightEqn}
\vect{w} = \mat{\Phi}^{-1}_g \vect{e}_{0} (\vect{e}_{0}^T \mat{\Phi}^{-1}_g \vect{e}_{0})^{-1}
\end{equation}

In conclusion, we see that a simple reinterpretation of Equation~\eqref{eqn:est_ang_freq} is all that is required to accommodate a two-component covariance model. The new weights $\vect{w}$ reflect the spatial covariance of residuals after the frequency mode subtraction rather than the spatial covariance of the sky itself.  In spirit, this is reminiscent of the approach in L13, where a best-guess model of the foregrounds was first subtracted from the sky maps before the variance of the residuals was minimized. The difference here is that our best-guess cleaning is based on the data rather than on a model. In addition, the estimator operates with separable spectral and spatial steps that are computationally trivial (a feature that we have managed to retain even with the \emph{non}separable two-component covariance of this section).

\subsection{Angular Weighting}
\label{ssec:angular_weighting}

Angular weighting is applied after the majority of the foreground covariance has been removed by the spectral mode subtraction $(1- \mat{F} \mat{F}^T)$. In the language of the previous section, it is not the full angular covariance $\mat{\Phi}$ of the foregrounds that acts as a weight, but instead just $\mat{\Phi}_g$, the angular covariance of the residuals. 

If instrumental systematics are well controlled, the eigenvalue spectrum of the $\nu-\nu'$ covariance will fall rapidly and settle to a thermal noise floor. Once the full complement of spectral modes down to the noise floor are removed, the map would then be dominated by thermal noise. The optimal angular weight in that case corresponds to dividing by the foreground template squared. Rather than suppressing any foregrounds, this operation simply weighs more strongly against regions of higher thermal noise due to brighter foregrounds that have boosted the $T_{\rm sys}$ (assuming foregrounds dominate the system temperature in these frequencies).

Our primary argument throughout this paper has been that foreground models should not be trusted to define spectral modes for foreground cleaning because they will invariably be too simple due to instrumental effects. In the case of angular weighting of thermal noise, though, the operation of dividing by the foreground template simply needs to be a best-guess thermal noise weight. Slight misspecification reduces the optimality of $\hat \alpha$ but does not introduce bias.

In experimental reality, the eigenvalue spectrum may fall more gradually and merge into the noise floor. This could be caused by the calibration, beam and polarization effects described in Section~\ref{sec:analysis_considerations}. As the eigenvalue falls to the thermal noise floor, the thermal noise spectral modes are strongly mixed with foreground modes, making the foregrounds difficult to cleanly separate. Further, Section~\ref{ssec:how_many_modes} describes how the number of foreground modes to remove may be ambiguous because of thermal noise. An $\ell$-by-$\ell$ weighting of the residuals then has the potential to improve the estimate of $\hat \alpha$ by down-weighting angular correlations that are consistent with residual foregrounds. Detecting some angular power spectrum in the $(1-\mat{F} \mat{F}^T)$-cleaned map that is measurably different from Gaussian thermal noise could indicate that a better weight $w_\ell$ is possible.

Given that the integration depth of an experiment is at the level of the cosmological signal (which is much smaller than the foregrounds), by definition most of the foreground spectral modes will be well measured. Deleting this set of spectral modes will remove the vast majority of foregrounds. Even if there are residual foregrounds mixed with the thermal noise, we expect that the optimality of the angular weight would not depend strongly on the assumed foreground spatial template.

%If the $\mat{F}$ modes cleanly separate the finite rank foregrounds, the spatial covariance $\mat{\Phi}_g$ that determines the angular weight represents Gaussian thermal noise. In this case $\mat{\Phi}_g$ diagonal, and proportional to $T_{\rm sys}^2$ based on the radiometer equation. Experimentally, synchrotron emission on the sky dominates the receiver noise temperature, so the angular weight $\vect{w}$ is simply to divide by the synchrotron template-squared.
%In reality, there will not be a clean separation of foregrounds and thermal noise. This is related to the fact that there is not a rigorous way to determine the number of modes to remove (Section~\ref{ssec:how_many_modes}). If the eigenvalue spectrum has a long tail, we can remove the obvious, high signal-to-noise modes, but some residual will remain. This means that the $\vect{w}$ needs to represent the angular covariance of the residual foregrounds. 

Following L13, we split the problem of nonstationarity and angular correlation of the angular covariance by writing 
\begin{equation}
\label{eq:mQm}
\mat{\Phi}_g = {\rm diag}(\vect{m}) \mat{Q} {\rm diag}(\vect{m})^T,
\end{equation}
where $\vect{m}$ is the foreground mean map. This is equivalent to dividing the foreground by the expected mean and then encoding the angular correlations of that normalized map in the matrix $\mat{Q}$. Let these correlations be full sky, isotropic, and diagonal. If $\mat{T}$ is a matrix that converts from real space to spherical harmonics, then the correlations will be diagonal $\mathbf{\Lambda} = \mat{T} \mat{Q} \mat{T}^T$, where $\mathbf{\Lambda}$ is just some function of $\ell$. The operation $\mathbf{\Phi}^{-1}$ can be interpreted algorithmically as 
\begin{itemize}
\item Divide each spatial slice of the data by a foreground model map (a synchrotron template).
\item Transform to spherical harmonic space and deweight $\ell$-by-$\ell$ by $w_\ell$ (defined below).
\item Transform back to map space and divide by the foreground model map again.
\end{itemize}

%Here the $\ell$ weighting reflects the angular correlations of any residuals after the $(1- \mat{F} \mat{F}^T)$ cleaning step. This is also the step where galactic masks could be applied in real space through a choice of $\vect{m}^{-1}$ that is zero in those regions.
%The choice of $\vect{w}$ is related to the optimality of the estimate of $\hat \alpha$. 
%In this regime, the thermal noise is not just a perturbation of bright foreground spectral eigenvalues. 
% Given the brightness of the foregrounds relative to the depth of integration, most of the foreground covariance should reside in, and be cleaned by a handful of modes. In this regime, the brightest spectral eigenmodes can still be cleanly removed, but the remaining covariance will be a mix of thermal noise and foregrounds.
%Note that in the limit that the residuals are purely thermal noise, $w_\ell = 1$, and dividing by $\vect{m}$ performs the $T_{\rm sys}$ weighting.

A simple, two-parameter model for the correlations is to let the residuals have some amplitude $\xi$ relative to thermal noise and correlation length $\sigma$, as
\begin{equation}
\label{eq:w_ell}
w_\ell = \left [ \xi e^{-\sigma^2 \ell (\ell+1) / 2} + e^{\theta_b \ell (\ell + 1)} \right ]^{-1}.
\end{equation}
Here we have assumed that the input data are deconvolved by the beam before applying angular weights (dividing by $B_\ell$). The term $e^{\theta_b \ell (\ell + 1)}$ describes thermal noise in that case. 
By picking the $\sigma$ and $\xi$ to give the lowest error, we may give preference to parameters where the error spuriously scatters low. A better alternative is to use the known instrumental thermal noise to inform $\xi$ and assume residuals in the data are correlated across the beam scale $\theta_b$ (not determining $w_l$ based on the error of $\hat \alpha$). Ideally, the residuals would be dominated by nearly Gaussian thermal noise, and $w_\ell=1$ would be a sufficiently optimal weight.

\begin{figure}
\centering 
\includegraphics[width=.45\textwidth]{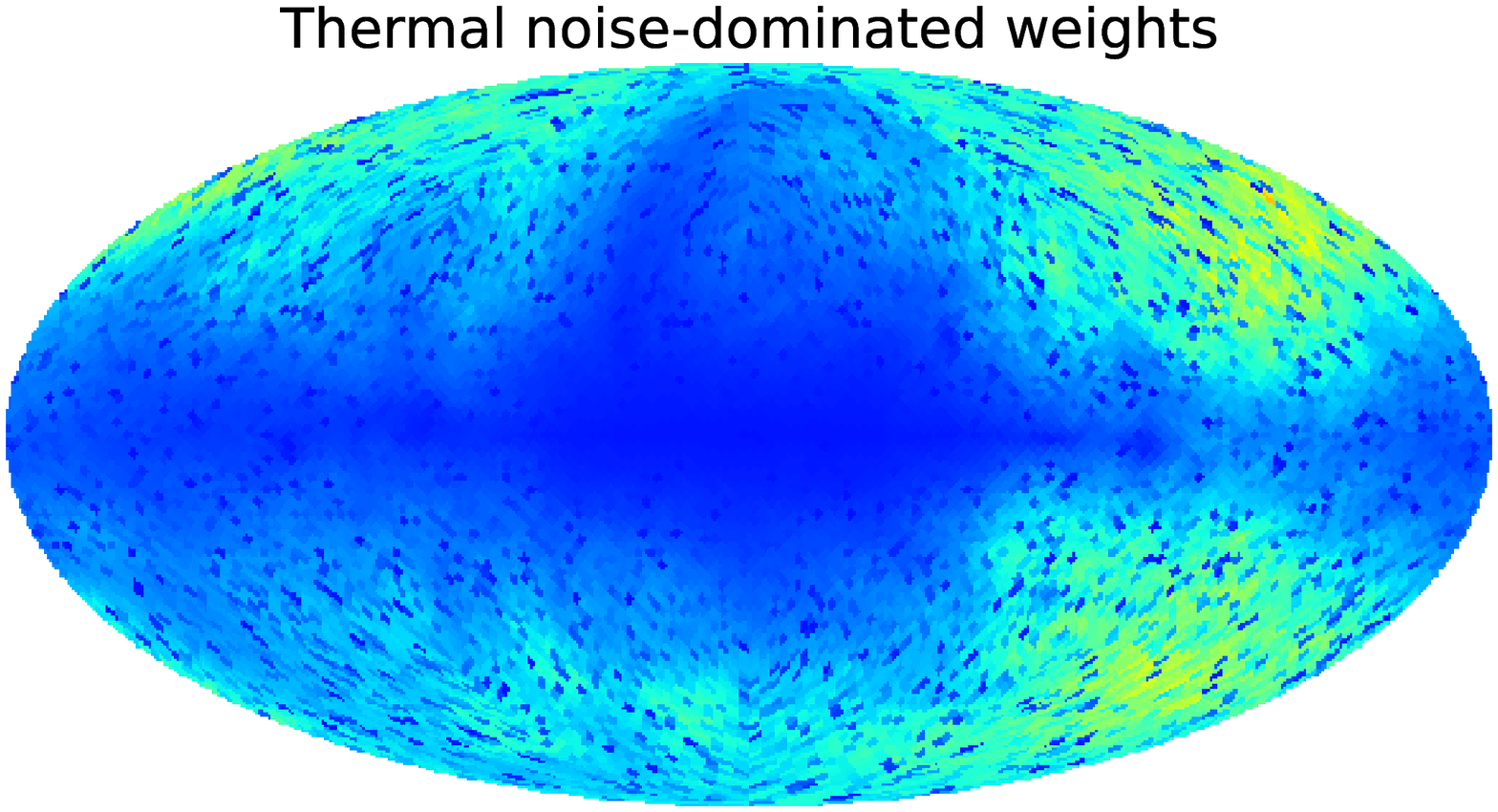}
\includegraphics[width=.45\textwidth]{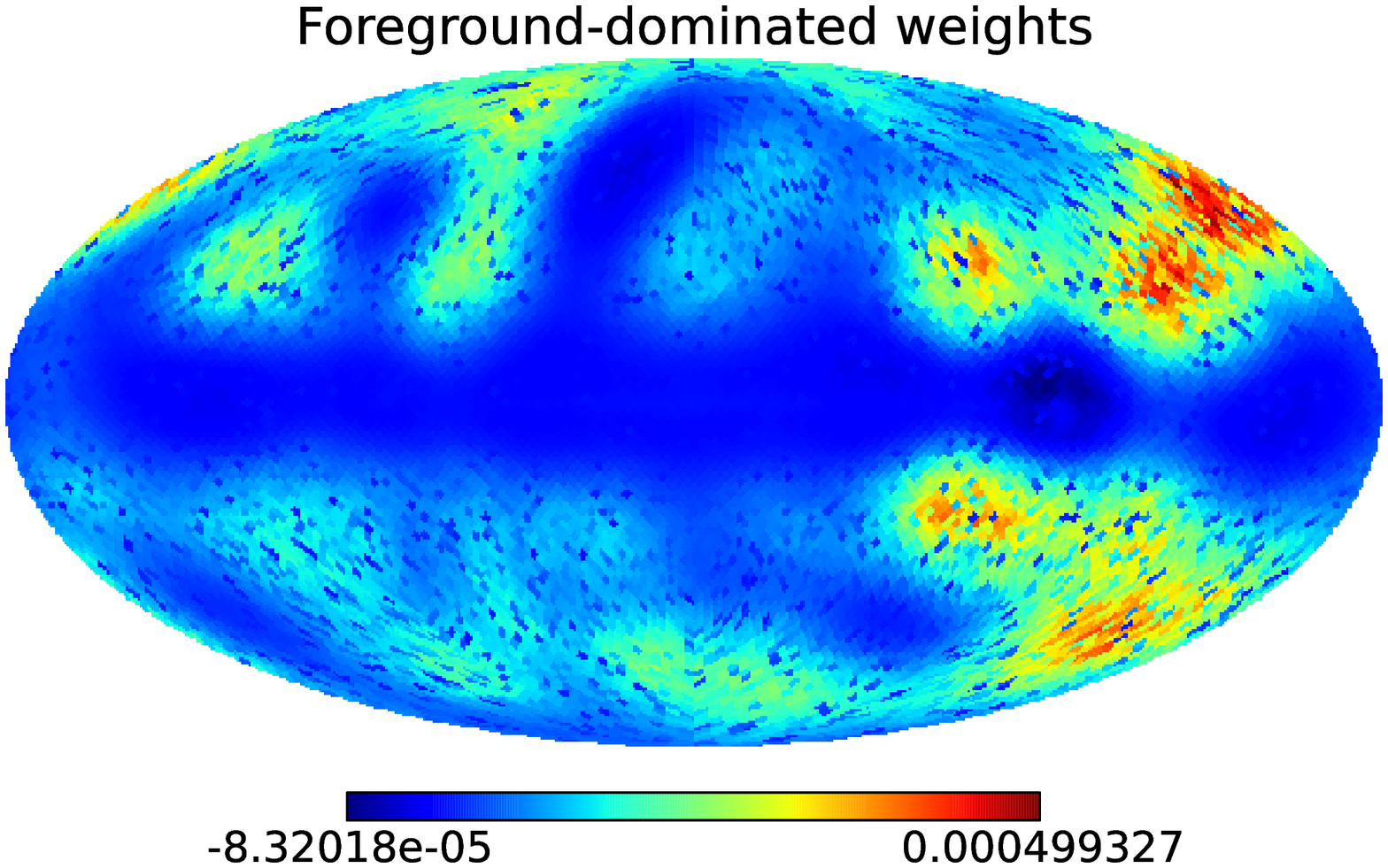}
\caption{{\it Top:} spatial weights at $80\,\textrm{MHz}$ for noise-dominated residuals ($\xi = 0.5$; $\sigma=20^\circ$).  The main purpose of spatial weighting in this regime is to average down noise, hence the weight is positive across the sky. Bright regions of the sky (which cause more instrumental noise for sky-noise-dominated instruments) are down-weighted.  {\it Bottom:} spatial weights at $80\,\textrm{MHz}$ for foreground-dominated residuals ($\xi = 50$; $\sigma=20^\circ$).  In this regime, the spatial weights play a role in foreground subtraction, and thus the weights are both positive and negative, as different parts of the sky are differenced to mitigate residual foregrounds. The beam size assumed in both cases is $\theta_b = 10^\circ$. The weights are defined to sum to one, but only the shape, not normalization, is relevant.
\label{fig:ExampleWeights}}
\end{figure}

In Figure~\ref{fig:ExampleWeights} we show example spatial weights computed from Equations~\eqref{eq:FinalWeightEqn}, \eqref{eq:mQm}, and \eqref{eq:w_ell} in two different regimes.  The top panel shows a noise-dominated case ($\xi=0.5$), and the bottom panel shows a foreground residual-dominated case ($\xi=50$).  In the noise-dominated case, the spatial weights are designed to simply average down the noise, and are therefore all positive.  The galactic plane and the bright point sources are seen to receive almost zero weight because they contribute the most thermal noise in a sky-noise-dominated instrument.  In the foreground-dominated case, the spatial weights attempt to further suppress residual foreground contamination.  There are thus both negative and positive weights because different parts of the sky can be differenced to subtract off foregrounds.  In both regimes, the right half of the sky is given more weight because the galaxy is dimmer there (see Figure~\ref{fig:fg_model}).

An important property of our spatial weighting scheme is that the normalization is irrelevant and only the shape matters. This is a particularly attractive property because most models of the sky at our frequencies of interest are based on interpolations and extrapolations from other frequencies. The amplitude may have large uncertainties, but the shape information is likely to be much more reliable. If one is particularly confident about the available shape information, it may be prudent to go one step further and set $w_{\ell = 0} = 0$ by hand. Examining Equations.~\eqref{eq:mQm} and \eqref{eq:w_ell} reveals that this projects out any component of the measured sky that has precisely the same spatial shape as our foreground model. Note that even though this is the $\ell=0$ weight, we do not destroy the global monopole signal that we seek to measure because the $w_\ell$ act in a prewhitened space \emph{after} dividing by $\vect{m}$.  The global signal therefore takes the form $1/\vect{m}$ and is no longer just the $\ell=0$ mode.  It is instead spread out over a wide range of $\ell$ values. Setting $w_{\ell = 0} = 0$ then sacrifices the sensitivity in one of the modes within this range, resulting in slightly increased error bars as one deviates slightly from the optimal prescription described above.  However, this may be a cost that is worth bearing for the sake of robustness in an aggressive campaign against foreground systematics.

\subsection{Summary of the Proposed Algorithm}
\label{ssec:algo_summary}

The final algorithm suggested here for analyzing global $21$\,cm data cubes is
\begin{itemize}
\item Use a spatially weighted average to project the monopole spectrum out of the map and find the $(\nu, \nu')$ sample covariance.
\item Find the largest eigenvectors of the sample covariance and project them out of the map.
\item Combine lines of sight using a prescription for the angular correlation of the residuals (which conservatively is to divide by the synchrotron template squared).
\item Dot this against the signal template to find the amplitude and perform error analysis.
\end{itemize}

While this algorithm is intuitive, our goal here has been to describe some of the implicit choices: (1) discovering frequency modes within the data (and implications for errors), (2) choosing to remove the part of the signal parallel to the foreground modes, and (3) assuming a two-component separable form for the thermal noise and foregrounds, respectively. The next section describes several considerations for using estimators of this type and challenges in global $21$\,cm signal measurement.

\section{Considerations for Analyzing Global $21$\, Maps}
\label{sec:analysis_considerations}

In the next few subsections, we will consider (1) the notion of resolution elements and the choice of beam size, (2) combined passband stability and resolution considerations, (3) challenges in error estimation from residual monopoles, (4) determination of the number of modes to remove, (5) extensions to simple amplitude constraints, (6) applications of our methods to the pre-reionization absorption feature, and (7) complications in foreground mitigation that result from Faraday rotation.

\subsection{Resolution Elements and Foreground Modes}
\label{sec:res_elt}

In Section~\ref{sec:simplifiedmodel}, the data were spectra of independent lines of sight. There, the number of foreground modes removed ($N_{\rm fg}$) compared to the number of sight lines ($N_\theta$) entered the errors as $\sim (N_{\rm fg} - N_\theta)^{-1}$. The sense of $N_\theta$ is more complex in our present case of an $N_\nu \times N_{\rm pix}$ $\mat{Y}$ data cube where the spatial part of the sky signal is convolved by an instrumental beam,. 

Intuition suggests that the number of independent sight lines is approximately the number of beam spots on the sky, but this is incorrect. If each spatial slice of a noiseless full-sky map is written in terms of its spherical harmonics $T_{\ell m}$, convolution by the beam is multiplication by $B_\ell$. Unless the beam falls to zero somewhere, this operation is invertible and so does not modify the rank of $\mathbf{\hat \Sigma}$. The rank of $\mat{Y}$ is $N_{\rm fg}$ so long as $N_{\rm pix} > N_{\rm fg}$ and $N_\nu > N_{\rm fg}$. Hence, all of the spectral modes can still be recovered from the noiseless beam-convolved map.

With wider beams, higher eigenvalue spectral eigenmodes (typically the less-smooth modes) are washed out by the beam.  These modes do not disappear entirely, but their amplitude diminishes.  We can develop a rough rule for this effect by finding the eigenvalue spectrum of full-rank white noise that is convolved by a given beam size. A fitting form for the suppression of eigenvalues (normalized to give no suppression for infinitely fine beams) is $b_n \approx 10^{-(\theta_{\rm FWHM} / 300^\circ)^2 f_{\rm sky}^{-1} n}$, as a function of eigenvalue $n$ and full width at half max (FWHM) of the beam.

At first sight, it would appear that a more rapidly falling eigenvalue spectrum is beneficial because it means that the foregrounds are better described by fewer covariance modes. When the suppression occurs because of beam convolution, however, it can be problematic in the presence of instrumental noise. With instrumental noise, the information about a mode may be suppressed below the noise floor. That foreground can therefore no longer be cleanly detected and removed and may contaminate the signal and lead to larger errors. Such detection and removal is potentially crucial because higher eigenvalues can still greatly exceed the signal (even after beam suppression) because of the hierarchy of scales.

We therefore come to the conclusion that even if angular resolution does not impose a hard limit on the number of foreground modes that may be discovered (as suggested in the case of independent sight lines), it does put a practical limit on the efficacy of the cleaning.  Note that the eigenvalues of $\mathbf{\Sigma} = \mathbf{\Sigma}_{\rm fg} + \mathbf{\Sigma}_{\rm inst}$ will be different from $\mathbf{\Sigma}_{\rm fg}$. Adding noise has the effect of mixing the high-variance foreground modes with the thermal noise modes, making it harder to cleanly separate foregrounds.

A rule of thumb for finding the number of modes discoverable before they are confused by noise is to plot the eigenvalue spectrum of noiseless foregrounds and the thermal noise, then find where they intersect. Modes some margin above this point will still be well measured. Consider a normalized eigenvalue spectrum of foregrounds that follows $\lambda_n \approx b_n \cdot 10^{-an}$ \citep{2012MNRAS.419.3491L}. The eigenvalues of thermal noise are relatively flat in $n$ in comparison. Finding the intersection with the noise level, the number of foreground modes that can be learned in the data is 
\begin{equation}
N_{\rm fg} \sim \log ({\rm FNR}) \left [ \left( \frac{\theta_{\rm FWHM}}{300^\circ} \right )^2 f_{\rm sky}^{-1} + a \right ]^{-1},
\end{equation}
where FNR is the foreground-to-noise ratio (in map variance). In our simple models here, $a$ may be a factor of a few, for only a few modes in total. In this idealized setting, the rather lenient scale of $300^\circ$ in our fitting form for $b_n$ means that an instrument could cover the full sky at poor resolution and still discover all of the foreground modes. In reality, instrumental effects will drive a higher number of required modes. For example, \citet{2013MNRAS.434L..46S} had to remove tens of modes to clean foregrounds even though only approximately four intrinsic synchrotron modes were expected. 

An advantageous factor here is that the sky signal is convolved by the beam whereas instrumental thermal noise is not. Hence, thermal noise can be estimated within the data using $N_{\rm pix}$ independent samples. In an ideal experiment, all of the foreground modes can be discovered and subtracted in $\mat{F}$ so that only thermal noise remains. Then errors could be assessed from the map's thermal noise, and signal loss could be corrected against the subtracted modes as $\vect{x}^T (1 - \mat{F} \mat{F}^T ) \vect{x}$.

It is generally beneficial to cover larger areas with finer resolution to better constrain and discover foreground modes. Higher resolution also helps focus deweighting of particular contaminated sky regions. An important exception to this occurs when the number of foreground modes scales with the FWHM or sky fraction. Both Faraday rotation and passband instability will generally increase the number of modes needed for studies on larger areas with smaller beams. Each new beam that is observed could see a new rotation measure or be observed with a different passband.

\subsection{Summarizing the Effects of Passband Stability and Angular Resolution}

\begin{figure}
\epsscale{1.1}
\plotone{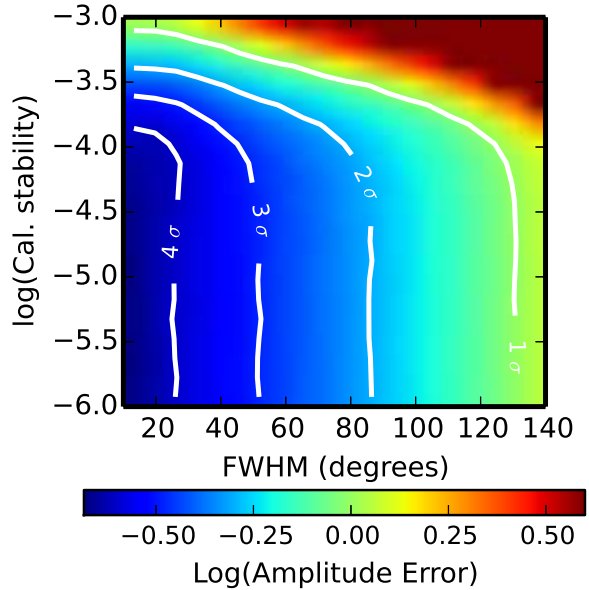}
\caption{Error of the recovered cosmological template amplitude as a function of calibration stability and instrumental resolution where the fiducial amplitude is $\alpha=1$. The thermal noise is set so that in the limit of no calibration or beam effects the amplitude can be constrained to $5 \sigma$. Here we consider foregrounds with four intrinsic spectral degrees of freedom. The constraint degrades for FWHM above $\sim 50^\circ$ and $10^{-4}$ stability in passband calibration. At low resolution, the modes cannot be discovered. Calibration stability worse than the thermal noise produces many more modes that must be constrained and removed. The Monte Carlo estimates in the upper right become very noisy, and we saturate the color scale.
\label{fig:error_dependence}}
\end{figure}

\begin{figure}
\epsscale{1.2}
\plotone{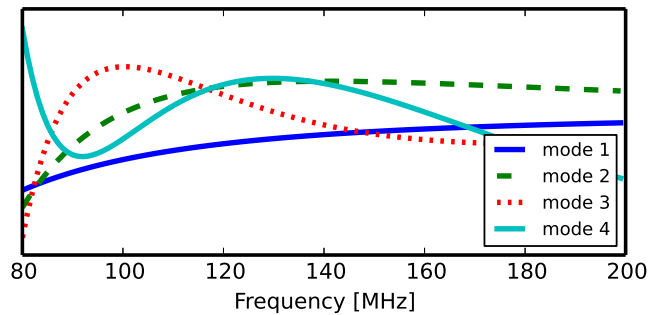}
\caption{Spectral eigenmodes of the foreground model described in Section~\ref{sec:globsig}. The second, third, and fourth modes are $3 \times 10^{-4}$, $8 \times 10^{-7}$, and $6 \times 10^{-8}$ down from the first mode (in variance), respectively.\label{fig:foreground_map_modes}}
\end{figure}

Figure~\ref{fig:error_dependence} summarizes two of the main points of our paper. Here, we use the galaxy model described in Section~\ref{sec:globsig} to perform a Monte Carlo simulation for the errors on $\alpha$ as a function of calibration stability and instrument resolution. Figure~\ref{fig:foreground_map_modes} shows the four largest modes of this extended GSM model. All further modes of the model itself (without noise) are negligible at the level of machine precision. Note that in contrast to the simple pedagogical models in Section~\ref{sec:simplifiedmodel} (with three modes), the full data cube simulation is more representative of the synchrotron sky. 

The sense of the passband calibration stability in our Monte Carlo is pixel to pixel (e.g., not beam convolved), so it is not affected by the beam convolution operation. We set the thermal noise level to produce a $5 \sigma$ detection of $\alpha$ in the absence of beam and calibration effects. For beams that are too large, not all foreground modes can be recovered and subtracted. In the simulations here with four modes, the constraint on $\alpha$  worsens gradually as the FWHM exceeds $\sim 50^\circ$. For calibration stabilities worse than $\sim 10^{-4}$, each line of sight responds differently to bright foregrounds, so each requires a new spectral function to remove. As soon as these passband calibration perturbations exceed the noise floor, they rapidly degrade the constraint.

Decreasing the thermal noise through longer integration times loosens the beam resolution requirement (because foreground modes are better measured). It does not modify the passband stability requirement. Passband stability is driven by the separation in scales between the foregrounds and the cosmological signal and the fact that each line of sight has a different response to the foregrounds. Even if the modes are well measured relative to thermal noise, they quickly overwhelm the signal. 

\begin{figure}
\epsscale{1.2}
\plotone{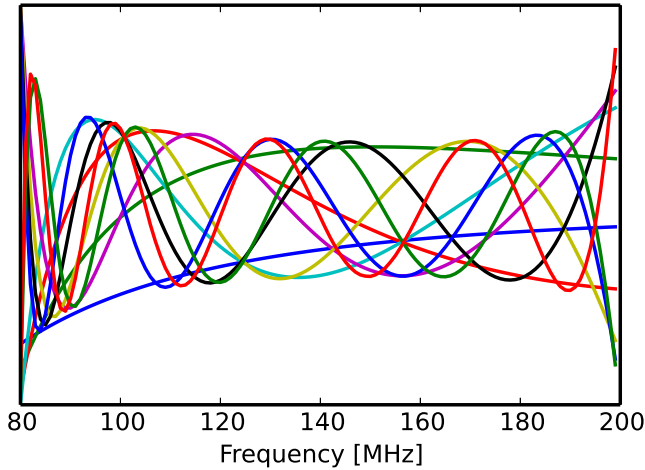}
\caption{Spectral eigenmodes of foregrounds in a diffraction-limited instrument where a Gaussian ${\rm FWHM} \propto \lambda$. Here, the beam's frequency dependence mixes spatial structure into spectral structure. This dramatically proliferates the number of spectral modes that are discovered and need to be removed, resulting in the loss of essentially all of the signal. A treatment for this effect in a mapping experiment is to convolve to a common resolution using the measured beams. Imperfections in the beam model will then generically produce additional modes of the spectral covariance. \label{fig:beam_modes}}
\end{figure}

Throughout, we have assumed that the angular resolution at all frequencies is constant. This is difficult to arrange experimentally. More typically, ${\rm FWHM} \propto \lambda$ by diffraction. In the frequency range considered here, a diffraction-limited FWHM would change by a factor of 2.5. Variation of the beam will mix spatial structure into frequency structure because different beams at each frequency see different angular contamination patterns \citep{2012ApJ...752..137M}. Implicitly, we have assumed that the beam is known well enough to convolve all of the frequency slices to a common resolution. In contrast, if no correction is performed, beam effects will proliferate the number of modes that need to be removed.  Figure~\ref{fig:beam_modes} shows the first $10$ bright spectral modes from the simulations described in Section~\ref{sec:globsig}, which originally had four modes. These new spectral modes describe the spatial structure that is mixed into the spectral direction. If these spectral modes had to be removed, most of the signal would also be projected out. The frequency-dependent beam must therefore be corrected to achieve efficient cleaning. Errors in the beam model executing this correction then translate into new spectral contamination modes. 

The tolerance of beam models is experiment-dependent and is beyond the scope of the present work but must be carefully considered. We emphasize that frequency-dependent beam effects impact any global $21$\,cm experiment, whether they choose to take advantage of angular information or not. The new spectral mixing modes reflect a best effort of our algorithm to describe a new spectral covariance that must be cleaned to reach the $21$\,cm monopole. Careful experimental design and measurement of the beam will control these covariance modes (which remain inaccessible in a single-beam measurement).

Here we have limited the intrinsic foreground rank to four. In reality, it may be discovered that foregrounds on the sky have higher rank, pushing the resolution constraint. Also, we assume a full sky here---an experiment covering a smaller fraction of the sky would need $f_{\rm sky}^{-1}$ more resolution in our simple approximations here.

\subsection{Challenges in Error Estimation}
\label{ssec:error_challenges}

In the absence of prior information, monopole contamination is indistinguishable from the signal. This contamination may arise from a synchrotron monopole or from an additive $T_{\rm rx}(\nu)$ of the receiver that is temporally constant but has some spectral structure. The foreground modes are estimated from the $(\nu, \nu')$ sample covariance $\mathbf{\hat \Sigma}$ of the map with the mean spectrum removed. The modes in $\mat{F}$ represent the frequency components of the spatially fluctuating part of the foregrounds, and the operation $1- \mat{F} \mat{F}^T$ is uninformed by a monopole signal.

The monopole foregrounds may have significant overlap with the spectral functions of the spatially fluctuating foregrounds. In our simulations here, the foregrounds are randomly drawn from a fixed set of spectral modes. For example, if there is a spatially varying synchrotron foreground with a spectral index of $-2$, that synchrotron emission will also be subtracted from the monopole by $1- \mat{F} \mat{F}^T$ . 

In reality, we cannot rule out the possibility of a foreground monopole that remains even after cleaning the spatially-varying synchrotron modes. Further, the spectral pattern of a constant instrumental $T_{\rm rx}(\nu)$ will never be discoverable by mapping across the sky and could be confused with the global signal. These present serious challenges for rigorous error estimation.

Performing the angular operations on the right of Equation~\eqref{eqn:est_ang_freq} first, the noise terms of $\hat \alpha$ can be isolated as
\begin{equation}
\hat \alpha = \alpha |_{\rm true} + \frac{\vect{x}^T (1- \mat{F} \mat{F}^T) }{\vect{x}^T (1- \mat{F} \mat{F}^T) \vect{x}} (\vect{n}_{\rm fg} + \vect{n}_{\rm inst})
\end{equation}
where $\vect{n}_{\rm fg}$ and $\vect{n}_{\rm inst}$ are the foregrounds and thermal noise of the monopole through $\mat{Y} \vect{w}$. We lump the constant $T_{\rm rx}(\nu)$ in with $\vect{n}_{\rm fg}$ rather than introducing a new term.

While the thermal noise contribution $\propto \vect{x}^T (1- \mat{F} \mat{F}^T)  \vect{n}_{\rm inst}$ has a well-defined distribution and can easily be included in the errors, $\vect{x}^T (1- \mat{F} \mat{F}^T)  \vect{n}_{\rm fg}$ is much more challenging. If $\vect{x}^T (1- \mat{F} \mat{F}^T)  \vect{n}_{\rm fg}$ were always positive, then $\hat \alpha$ could be interpreted as an upper bound. This practice is common in treating foregrounds to the power spectrum because their contribution is always positive in the quadratic quantity \cite{2013MNRAS.434L..46S, 2014ApJ...788..106P, 2014PhRvD..89b3002D, 2013MNRAS.433..639P}. Here, we have no guarantee that the residual foreground monopole dotted into the signal is positive. Recall that even though synchrotron radiation may be a strong power law, the residuals here will largely represent unknown instrumental factors that could dot with an arbitrary sign with the signal template. It has long been known that higher-order polynomials lead to better subtractions. This suggests that residuals will generally produce two-sided errors. Further, any monopole residual foregrounds are likely to be non-Gaussian, inheriting from the non-Gaussianity of the bright foregrounds.

These ``monopole'' foregrounds do not strictly need to be a monopole on the sky---they only have to be constant over the area of the survey, if that only covers part of the sky. This again bolsters the argument that global $21$\,cm experiments should observe as much of the sky as possible with modest resolution to maximize their sensitivity to spatial variations that could lead to significantly cleaner signal recovery.

Barring any prior information about the foregrounds or instrumental response, the best ways to immunize the analysis against this response are (1) to document that the spectral cube cleaned with $(1- \mat{F} \mat{F}^T)$ is consistent with Gaussian thermal noise and (2) to cover a large area of the sky with modest resolution to capture additional spatial variation of foregrounds. Additional information about spectral smoothness and instrument response can be used \citep{2010Natur.468..796B, 2014ApJ...782L...9V} to further clean the spectral monopole. Quoted errors would be based only on thermal noise, and boosted to account for signal lost in the cleaning process, Equation~\eqref{eqn:gaussian_mode_error}. Another possibility would be to conduct, for example, surveys of the north and south celestial poles with different receiver architectures. If these were performed independently and had the same result up to thermal errors, it could lend some confidence that any residual monopoles in the respective surveys are small compared to noise.

\subsection{Determining the Number of Modes to Remove}
\label{ssec:how_many_modes}

\begin{figure}
\epsscale{1.2}
\plotone{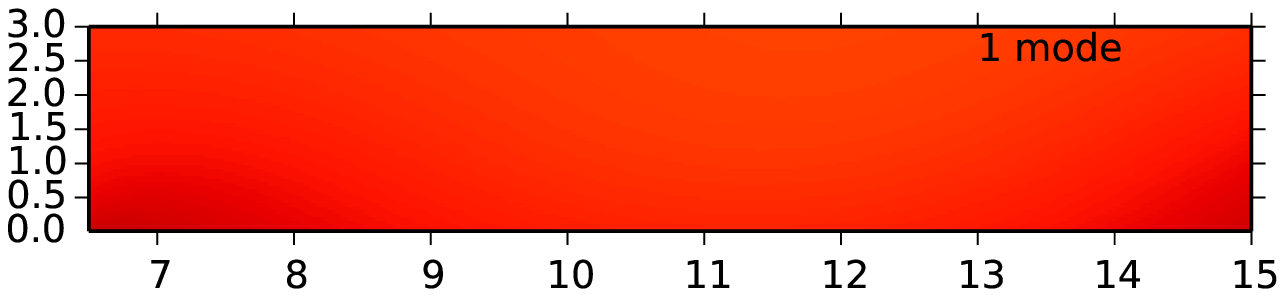}
\plotone{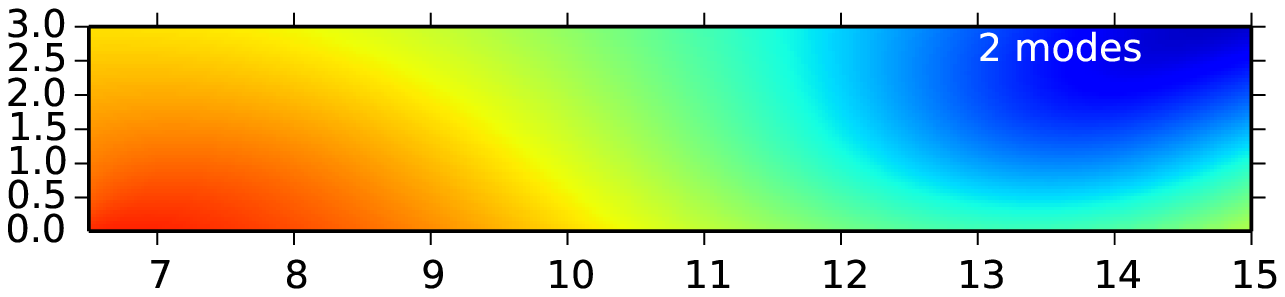}
\plotone{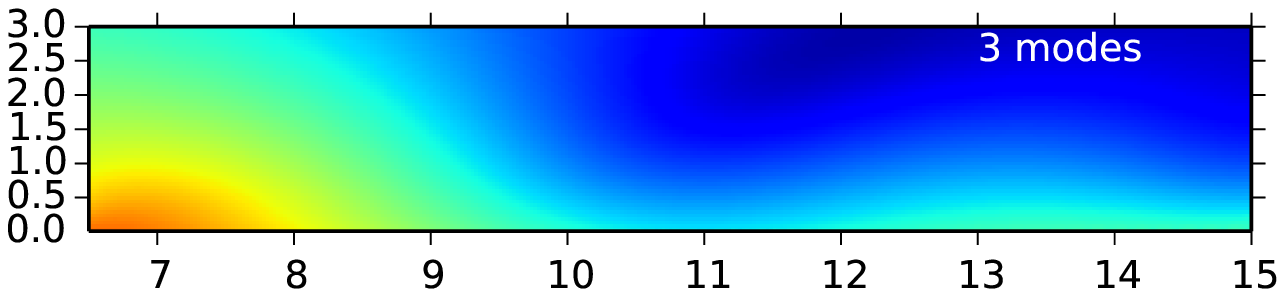}
\plotone{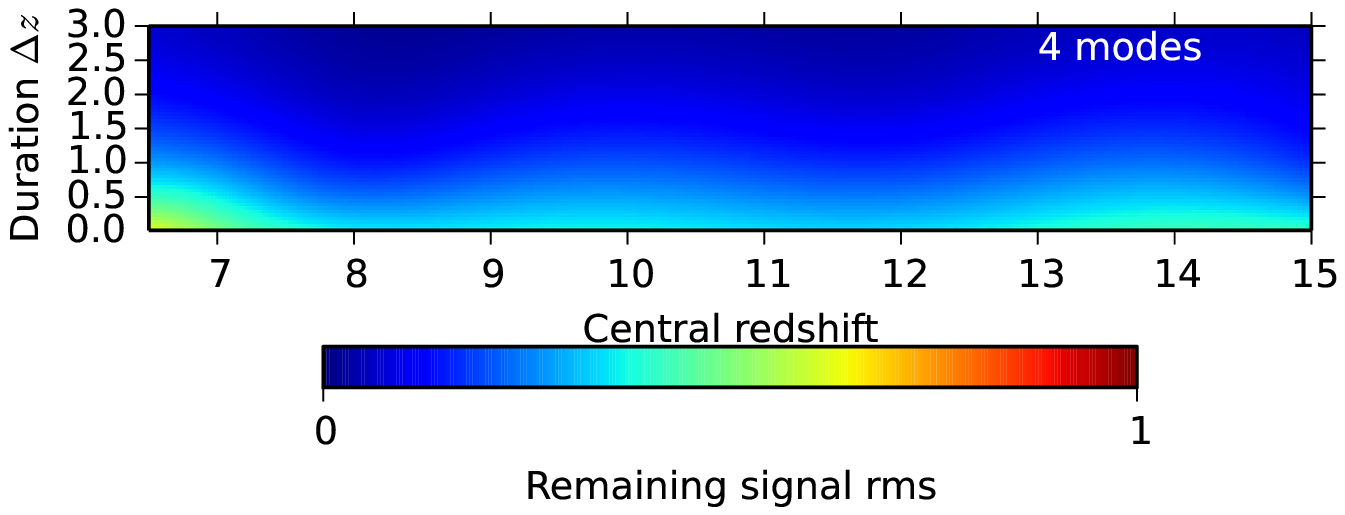}
\caption{Fraction of the root mean squared of the reionization signal remaining after the first four foreground modes are removed. The signal is given by Equation~\eqref{eqn:tanh_model} for a simple reionization scenario with central redshift $z_r$ and width $\Delta z$. The first mode is a power law (see Figure~\ref{fig:foreground_map_modes}), and most of the signal remains for these reionization scenarios. The other three modes carry more redshift information and tend to express signal variations at high redshift better. In all cases, as the signal becomes smoother ($\Delta z$ increasing), it becomes harder to differentiate from the foreground modes.\label{fig:remaining_signal}}
\end{figure}

Figure~\ref{fig:remaining_signal} shows the fractional rms of the cosmological reionization signal remaining after removing the four foreground modes of the extended GSM (Section~\ref{sec:globsig}), shown in Figure~\ref{fig:foreground_map_modes}. We use the operation $(1- \mat{F} \mat{F}^T)$ described previously. The foreground spectral functions overlap strongly with the signal, especially for slower, spectrally smoother reionization histories (high $\Delta z$ in the plot). Nulling four foregrounds causes significant signal loss over a range of reionization models. Any instrumental response that increases the rank of these foregrounds will result in additional signal loss from the new modes it introduces.

In the case of these simulations, we know the precise number of spectral degrees of freedom. In experimental practice, the number of contaminated modes cannot be determined directly. This can be made especially ambiguous by an eigenvalue spectrum that drops slowly rather than reaching a clear noise floor.

A central challenge of any global $21\,\textrm{cm}$ signal estimator is determining the number of modes to remove \citep{rao1967, KolloRosen}. In similar methods applied to the $21$\,cm autopower \citep{2013MNRAS.434L..46S}, an advantage is that the bias from foregrounds is purely additive. In this case, an experiment can report the constraint with errors as a function of modes removed. This will generically fall as more contaminated modes are removed. This decline will level out if the foreground have been discovered. As further modes are removed, the errors will increase due to cosmological signal loss, which is self-consistently included in our formalism. A reasonable prescription for a bound is to find the number of modes that gives the most stringent upper bound.

In the global spectrum case, foreground residuals after $N$ spectral modes are removed can have either sign in the dot product with the signal template. This two-sided bias should stabilize as most of the foreground modes are discovered. For more modes removed, the errors will grow, but the central value should remain stable if most of the foreground covariance is described and removed. The summary results of a global $21$\,cm experiment should include plots of the spectral modes removed and characterization of the constraint as a function of modes removed.

Both this issue and the residual monopole in Section~\ref{ssec:error_challenges} cannot be addressed rigorously. An irreducible challenge of global $21$\,cm measurement is determining whether any foreground bias remains. The methods here provide excellent prospects for cleaning the data and guiding an experiment, but they do not solve these central issues.

\subsection{Extending the Spectral Template}
\label{ssec:extended_template}

We have discussed constraints on the amplitude of a cosmological $21$\,cm signal template that is known in advance. The primary goal of the current generation of instruments is discovery of this amplitude. Subsequent experiments will then constrain the spectrum itself. The methods developed here can also be applied to that case. 

There is a continuum of models between the template estimate here and a full spectrum. One possibility would be to use prior information about the redshift of reionization and estimate two amplitudes, $\alpha (z < z_r)$ and $\alpha (z > z_r)$, and then examine the significance of a step feature. Another option would be to constrain a fixed number of spectral spline amplitudes. In the limit that each frequency bin has a different amplitude, the methods developed here can find the component of that signal orthogonal to the identified foreground modes.

The choice between models must juggle (1) use of prior information, (2) discriminatory power, and (3) insight about the trajectory of the global $21$\,cm signal. This model selection problem is deferred to future work.

\subsection{Application to Pre-reionization Absorption}
\label{sec:DarkAges}

\begin{figure}
\epsscale{1.2}
\plotone{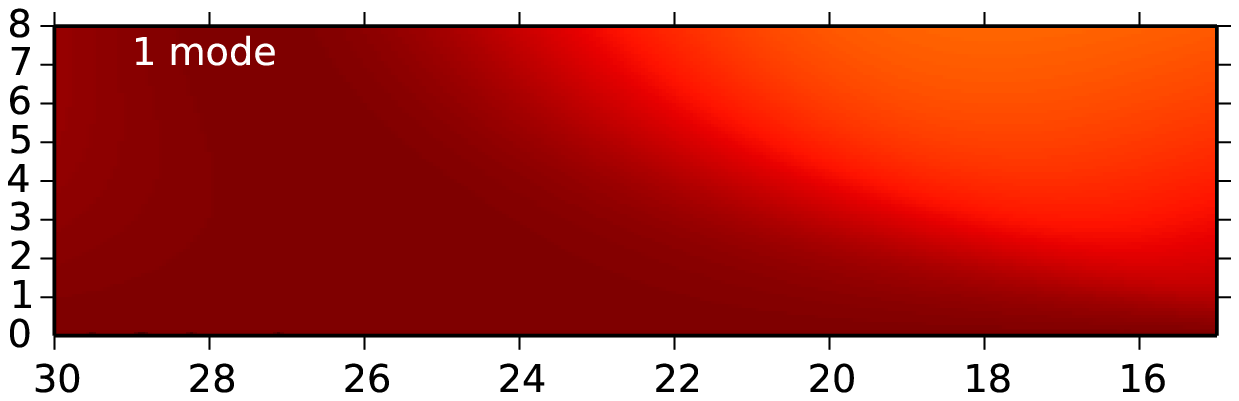}
\plotone{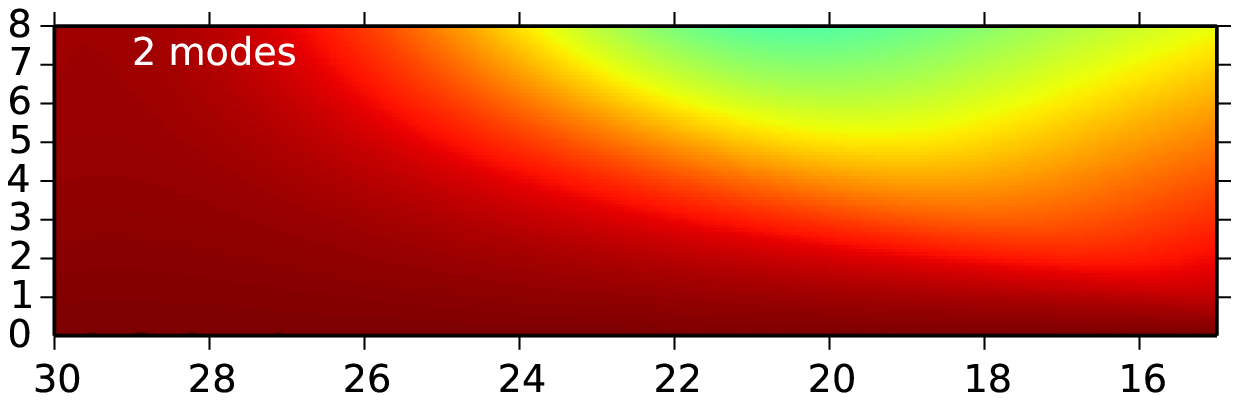}
\plotone{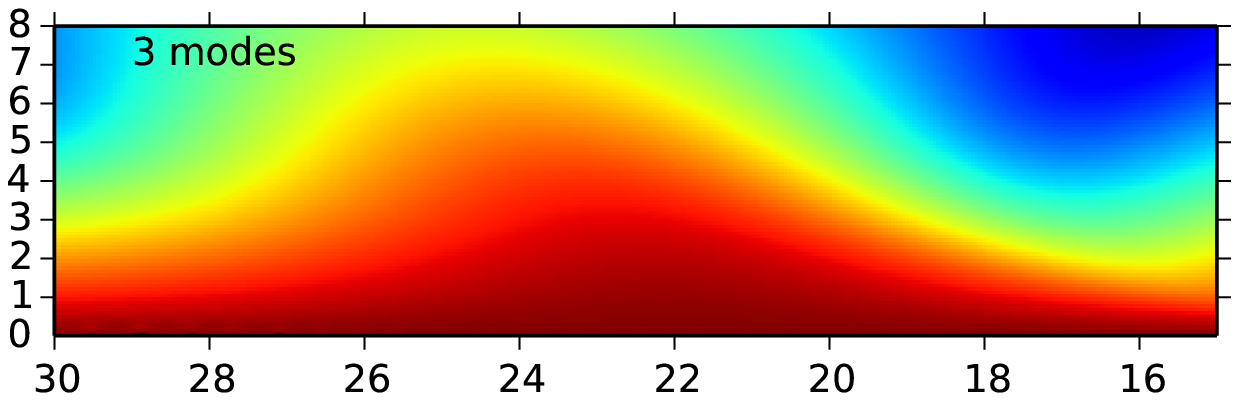}
\plotone{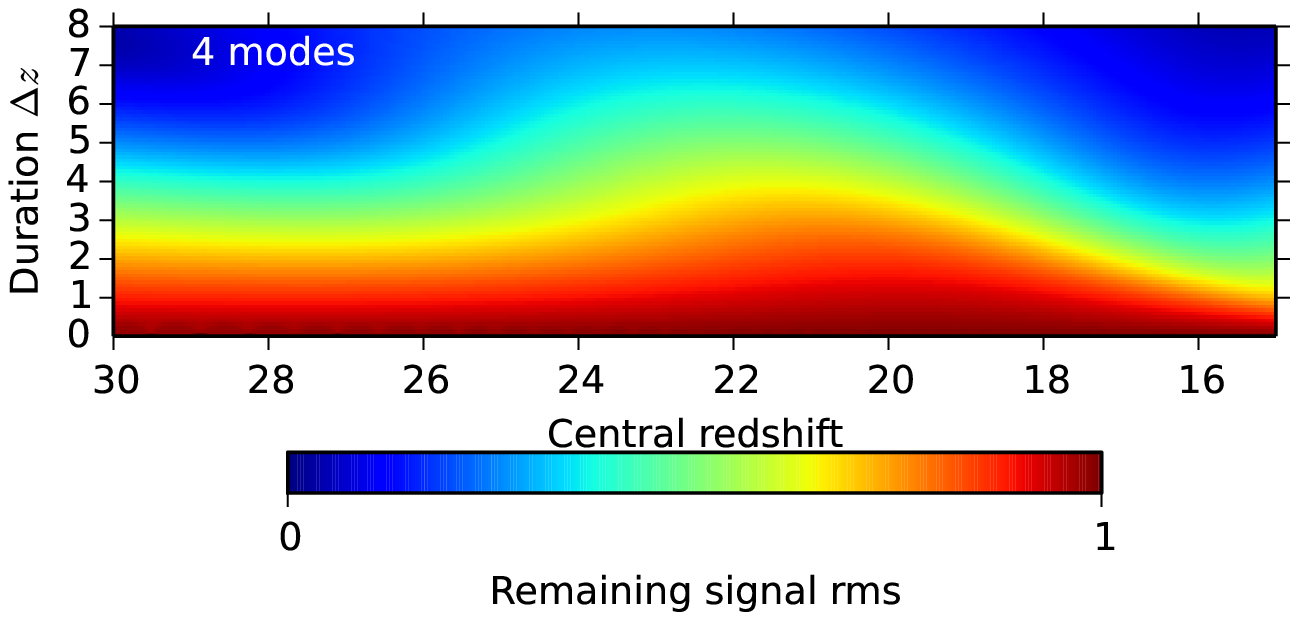}
\caption{Same as Figure~\ref{fig:remaining_signal}, except that we now consider a pre-reionization absorption feature parameterized by a Gaussian. The duration is given as the $\sigma$ width of the Gaussian. Note that significantly more signal remains.\label{fig:remaining_signal_dark}}
\end{figure}

Throughout the paper, we have considered the global $21$\,cm signal from reionization. This choice was primarily pedagogical, to have a simple step around which the foregrounds are subtracted. The maximum contrast of this feature is $\sim 30$\,mK, and for even modest $\Delta z$ it becomes smooth and more easily subtracted by the foreground modes. In contrast, in the period prior to reionization, the $21$\,cm is thought to go into absorption at $\sim 100$\,mK \citep{2010PhRvD..82b3006P}. Here, the spin temperature is strongly radiatively coupled to the cold gas. However, while synchrotron radiation contaminating $z \sim 10$ $21$\,cm radiation around the time of reionization is $\sim 500$~K (averaged across the sky), the contamination of an absorption signature at $z \sim 20$ is $\sim 3000$~K. 

While the synchrotron emission from higher redshift $21$\,cm emission may be greater, the absorption signature has the additional feature that it is two-sided, both falling and rising. This generally makes foreground modes less parallel to the signal and leads to less signal loss. For our purposes here, we can again develop a simple model where $T_b$ is a Gaussian function parameterized by a central redshift and width $\sigma_z$. Figure~\ref{fig:remaining_signal_dark} is analogous to Figure~\ref{fig:remaining_signal} except that it considers this era. Generally much more cosmological signal remains orthogonal to the foreground modes.

\subsection{Susceptibility to Faraday Rotation}
\label{ssec:faraday}

In addition to the Stokes-$I$ component of the foregrounds, synchrotron emission is also polarized. The polarized emission is subject to Faraday rotation as it traverses various column densities of free electrons in magnetic fields. If a receiver were designed to be purely sensitive to Stokes-$I$, then this spectral fluctuation could be ignored. Generally though, a receiver will have some level of response to polarized signals, and Faraday rotation produces a signal that oscillates in frequency \citep{2013ApJ...769..154M}. The rotation angle is $({\rm RM}) \cdot \lambda^2$, where $\textrm{RM}$ is the rotation measure.  Even for modest RM, the rotation angle can vary rapidly over our band. This is especially problematic for the global $21$\,cm signal because the rotation measure varies over the sky, increasing the number of contaminated modes in the data.

\begin{figure}
\epsscale{1.2}
\plotone{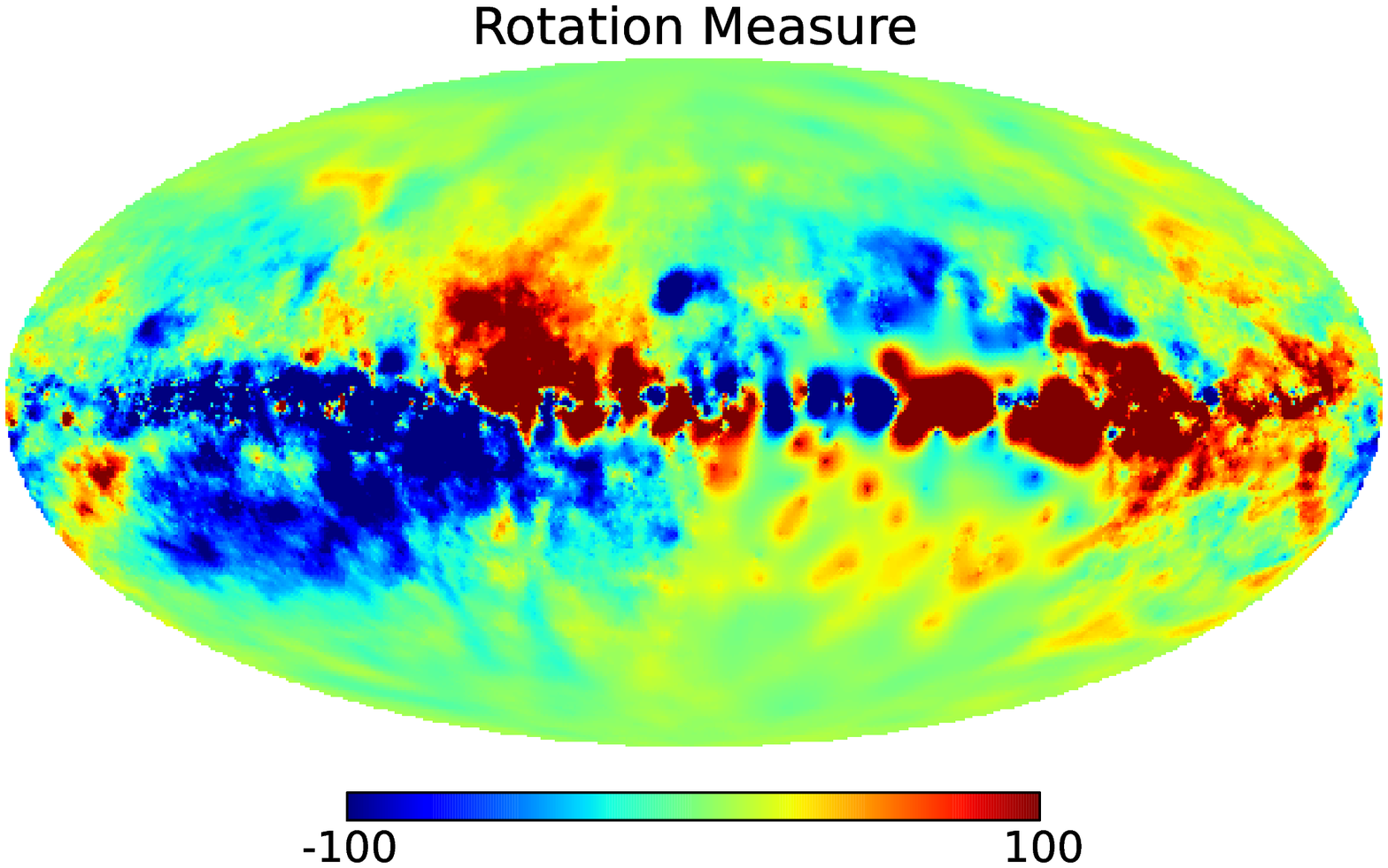}
\plotone{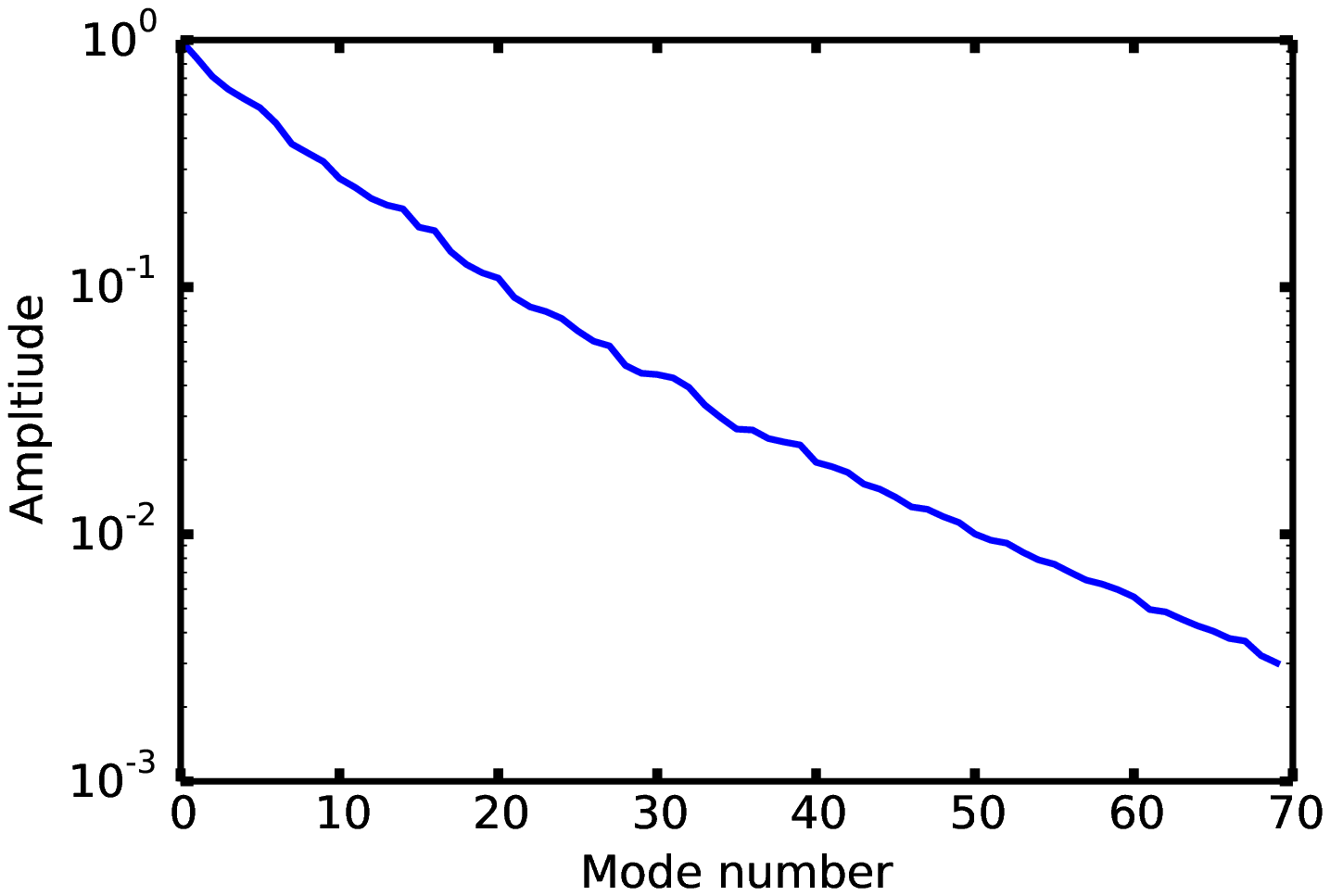}
\caption{{\it Top:} galactic rotation measure map of \citet{2014arXiv1404.3701O} in units of ${\rm rad}\,{\rm m}^{-2}$, with color range truncated at $\pm 100~{\rm rad}\,{\rm m}^{-2}$ to show more structure out of the plane. {\it Bottom:} the eigenvalue spectrum of Stokes-$Q$ emission in the galactic model of \citet{2009A&A...495..697W}, normalized to one. This component falls off very slowly because the rotation measure introduces many degrees of freedom that vary across the sky.
\label{fig:polarized_modes}}
\end{figure}

Extragalactic sources are subject to Faraday rotation through the screen of the entire Milky Way. This has been measured recently by \citet{2012A&A...542A..93O, 2014arXiv1404.3701O}, and is shown in Figure~\ref{fig:polarized_modes}. Each point source will generally have a different rotation measure and require a new degree of freedom to describe. \citet{2013ApJ...771..105B} has conducted deep, wide-field observations of polarization of galactic synchrotron emission at $189$\,MHz. They find only one source out of 70 with $S>4$~Jy with polarization at the level of $1.8\%$ in $2400\,{\rm deg}^2$, with all others falling below the $2\%$ polarization fraction. 

Polarization and Faraday rotation of galactic synchrotron emission is more complex because synchrotron emission is interspersed with the Faraday screen. \citet{2013ApJ...771..105B} find peak polarized emission at $\sim 13$~K and rotation measures mostly below $10~{\rm rad}\,{\rm m}^{-2}$. The low rotation measures and polarization fraction are thought to be due to a depolarization horizon \citep{2010A&A...520A..80L, 2003ApJ...585..785U} within the galaxy, a distance beyond which most of the emission is depolarized along the line of sight. 

The global $21$\,cm experiments to date have been single element, making it difficult to make a precise comparison with interferometric galaxy polarization surveys conducted at similar frequencies \citep{2009MNRAS.399..181P, 2013ApJ...771..105B}. \citet{1996AstL...22..582V} measured $3.5 \pm 1.0$~K at $88$\,MHz ($24^\circ$\,FWHM, phased array) and $2.15 \pm 0.25$~K at $200$\,MHz ($8^\circ$\,FWHM, single dish) for the value of $\sqrt{Q^2 + U^2}$ toward the north celestial pole. 

To assess the galactic polarization field, we use the model of \citet{2009A&A...495..697W}, which self-consistently simulates the polarized emission and rotation measure. Figure~\ref{fig:polarized_modes} shows the eigenvalue spectrum of the Stokes-$Q$ spectral modes. Unlike Stokes-$I$, polarization is spectrally much more complex, and its covariance is spread among many modes. Ideally, all of these modes would be suppressed well below the noise level of the experiment so they do not need to be estimated and subtracted. Based on the measured amplitudes of the polarized signals above, suppose that polarization is $10^{-3}$ of intensity. To achieve a $10^5$ suppression, the polarization leakage must be kept below $1\%$.

\section{Conclusion}
\label{sec:discussion}

Measurements of the global $21$\,cm signal are very challenging due to (1) astrophysical foregrounds and their interaction with instrumental systematics, (2) terrestrial radio interference, and (3) the ionosphere. Here we have examined the first issue, especially in regard to instrumental response. Following $z\sim 1$ $21$\,cm literature \citep{2013MNRAS.434L..46S, 2013ApJ...763L..20M, 2010Natur.466..463C}, we develop a new method where foregrounds can be jointly estimated with the monopole spectrum. This relies on the fact that foregrounds vary across the sky while the cosmological signal is constant. This idea also extends L13, who argue that surveys with moderate angular resolution covering much of the sky are able to better discriminate between the monopole $21$\,cm signal and foregrounds.

The key observation arising from our cleaning method is that the instrument should be designed to minimize the generation of new spectral degrees of freedom from the foregrounds. For example, if each line of sight has a slightly different passband calibration, it also requires a new spectral degree of freedom to describe the foregrounds there. In this sense, the instrument ``spreads out'' the variance over modes that ultimately require more aggressive cleaning and signal loss. In contrast, a constant passband calibration error does not increase the rank of the foreground spectral covariance, and its effect is primarily aesthetic. This is fortuitous because obtaining a smooth spectral calibration of an instrument at these frequencies would require very large, expensive structures that are black in radio wavelengths. Simply requiring that galactic foregrounds be smooth to a few percent provides a sufficient passband calibration for signal recovery. In contrast, significant effort must be put into maintaining passband stability to at least $\sim 10^{-4}$. This may require using much of the instrument's sensitivity to integrate against a stable reference.

Polarization to intensity leakage is another example of an instrumental systematic that increases the rank of the foreground covariance. It allows spectral oscillations in the Faraday-rotating polarization signal to contaminate the spectral intensity measurement. Because of depolarization, sky polarization is only a fraction of the intensity, and our estimates of the instrumental constraints are more lax, $< 10^{-2}$. 

In attempting to measure the global $21\,\textrm{cm}$ signal, one is faced with bright contaminating foregrounds that can interact with instrumental systematics in nontrivial ways, overwhelming the cosmological signal that one seeks to measure.  In this paper, we have developed methods that bear similarities to various previously proposed intuitive data-analysis techniques.  However, our methods arise from a rigorous, self-consistent framework that allows unknown and unanticipated foreground properties to be derived from real data.  Such a framework also provides guidance for instrument design.  If design requirements can be adequately met, high-significance measurements of the global $21\,\textrm{cm}$ signal will be possible, providing direct access to the rich and complex astrophysics of the first luminous objects and reionization.

\acknowledgements 
E.S. thanks Ue-Li Pen for conversations and feedback, and for stimulating the approach in collaboration on $z\sim 1$ $21$\,cm data analysis from the Green Bank Telescope. A.L. thanks Matt McQuinn for conversations. We thank Harish Vedantham for comments and the Ohio $21$\,cm workshop for fostering conversations that initiated the project. A.L. acknowledges support from NSF grants AST-0804508, AST-1129258, and AST-1125558.

\appendix

\section{A. Relation Between Covariance Adjustment and the Maximum-Likelihood Linear Estimator}
\label{app:lemma_7b}

In Section~\ref{sec:cov_from_meas}, we claim that the estimator $\vect{m} = (\mat{X}^T \mathbf{\Sigma}^{-1} \mat{X})^{-1} \mat{X}^T \mathbf{\Sigma}^{-1} \vect{x}$ is equivalent to the projection operation, $\vect{m} = (\mat{X}^T \mat{X})^{-1} \mat{X}^T (1 - \mathbf{\Pi}) \vect{y}$ where $\mathbf{\Pi} = \mathbf{\Sigma} \mat{Z} (\mat{Z}^T \mathbf{\Sigma} \mat{Z})^{-1} \mat{Z}^T$.  The two essential conditions here are that $\mat{X}^T \mat{Z} = 0$ (that the signal and nonsignal vectors are orthogonal) and that the vectors in $\mat{X}$ and $\mat{Z}$ together span the space. The operations $\mat{X}^T$ and $\mat{Z}^T$ can be understood as projecting onto the signal basis and the everything-but-signal basis, respectively. This proof follows Lemma 2b of \citet{rao1967}. Writing out the terms, we would like to prove
\begin{equation}
(\mat{X}^T \mathbf{\Sigma}^{-1} \mat{X})^{-1} \mat{X}^T \mathbf{\Sigma}^{-1} = (\mat{X}^T \mat{X})^{-1} \mat{X}^T - (\mat{X}^T \mat{X})^{-1} \mat{X}^T \mathbf{\Sigma} \mat{Z} (\mat{Z}^T \mathbf{\Sigma} \mat{Z})^{-1} \mat{Z}^T.
\end{equation}
Multiplying on the right by $\mat{X}$ is trivially true because $\mat{Z}^T \mat{X} = 0$ and the other terms are the identity matrix. This checks just the $\mat{X}$ subspace of the equality. To prove general equality we need to check the $\mat{Z}$ subspace. Multiply on the right by $\mat{Z}$ and from the left by $\mat{X}^T \mathbf{\Sigma}^{-1} \mat{X}$ (which will allow simplification)
\begin{eqnarray}
\mat{X}^T \mathbf{\Sigma}^{-1} \mat{Z} = -(\mat{X}^T \mathbf{\Sigma}^{-1} \mat{X}) (\mat{X}^T \mat{X})^{-1} \mat{X}^T \mathbf{\Sigma} \mat{Z} (\mat{Z}^T \mathbf{\Sigma} \mat{Z})^{-1} \mat{Z}^T \mat{Z} \\
\mat{X}^T \mathbf{\Sigma}^{-1} \mat{Z} = -(\mat{X}^T \mathbf{\Sigma}^{-1} \mat{X}) (\mat{X}^T \mat{X})^{-1} \mat{X}^T \mathbf{\Sigma} \mat{Z} [ (\mat{Z}^T \mat{Z})^{-1} (\mat{Z}^T \mathbf{\Sigma} \mat{Z})]^{-1} \\
\mat{X}^T \mathbf{\Sigma}^{-1} \mat{Z} [ (\mat{Z}^T \mat{Z})^{-1} (\mat{Z}^T \mathbf{\Sigma} \mat{Z})] = -(\mat{X}^T \mathbf{\Sigma}^{-1} \mat{X}) (\mat{X}^T \mat{X})^{-1} \mat{X}^T \mathbf{\Sigma} \mat{Z} \\
\mat{X}^T \mathbf{\Sigma}^{-1} [ \mat{Z} (\mat{Z}^T \mat{Z})^{-1} \mat{Z}^T + \mat{X} (\mat{X}^T \mat{X})^{-1} \mat{X}^T ] \mathbf{\Sigma} \mat{Z} = 0.
\end{eqnarray}
Because $\mat{X}$ and $\mat{Z}$ span the space, $\mat{Z} (\mat{Z}^T \mat{Z})^{-1} \mat{Z}^T + \mat{X} (\mat{X}^T \mat{X})^{-1} \mat{X}^T = \mat{1}$, then noting that $\mat{X}^T \mat{Z} = 0$ proves the identity. In many of the methods in the paper, the projection $\mathbf{\Pi}$ is reduced to $(1- \mat{F} \mat{F}^T)$, completely projecting out a set of modes. The modes in $\mat{F}$ clearly cannot span the space. Further, we let the modes $\mat{F}$ overlap with the signal. These choices make the estimator less formally optimal but more robust to very bright, non-Gaussian foregrounds.

\section{B. Extending the Monopole Constraint: the Growth Model}
\label{sec:growth_curve}

The separable model for the monopole can be extended to a more general spatial-spectral template scheme. Rather than a simple signal model $\alpha \vect{x} \vect{e}_0^T$, the mean of the observed data can be described by the outer product spectral modes $\vect{u}_i$ and spatial modes $\vect{v}_i$ (not necessarily mutually orthogonal). In terms of the $\nfreq \times \npix$ data cube $\mat{Y}$, 
\begin{equation}
\mat{Y} = \mat{U} \mat{X} \mat{V}^T = \sum_{i,j} x_{ij} \vect{u}_i \vect{v}_j^T.
\end{equation}
This extended model could represent spatial and spectral templates of galactic emission in addition to the monopole signal. The noise terms in the model would then describe any residuals with respect to this model, written as 
\begin{equation}
\mat{Y} = \mat{U} \mat{X} \mat{V}^T + \mathbf{\Sigma}^{1/2} \mat{E} \mathbf{\Phi}^{1/2}
\end{equation}

This separable estimation problem is considered in detail by \citet{KolloRosen}. (To reach their notation, right multiply by $\mathbf{\Phi}^{-1/2}$.) Form the weighted $(\nu, \nu')$ sample covariance with the spatial components of the signal projected out, as 
\begin{equation}
\mat{S} = \mat{Y} \mat{\Phi}^{-1} (\mat{1} - \mat{V} \mat{W}_{\mat{V}}) \mat{Y}^T
\end{equation}
where $\mat{W}_{\mat{V}} = (\mat{V}^T \mat{\Phi}^{-1} \mat{V})^{-1} \mat{V}^T \mat{\Phi}^{-1}$ is the standard maximum-likelihood estimator for the amplitudes of the spatial vectors in $\mat{V}$. The maximum-likelihood estimator for $\mat{X}$ and full-rank spectral foregrounds $\mat{S}$ is 
\begin{equation}
\mat{\hat X} = (\mat{U}^T \mat{S}^{-1} \mat{U})^{-1} \mat{U}^T \mat{S}^{-1} \mat{Y} \mat{W}_{\mat{V}}^T.
\label{eqn:fullrankseparable}
\end{equation}
\citet{KolloRosen} describe extensions of this more general estimation problem to finite rank $\mat{S}$.

\bibliographystyle{apj}
\bibliography{global_21cm}

\end{document}